\documentclass{article}
\usepackage{latexsym}

\begin{document}

\begin{center}
IMPROVED LABORATORY TRANSITION PROBABILITIES FOR Sm \textsc{ii} AND 
APPLICATION TO THE SAMARIUM ABUNDANCES OF THE SUN AND THREE $r-$PROCESS RICH, 
METAL-POOR STARS
\end{center}

\begin{center}
(short title: Sm Transition Probabilities and Abundances)
\end{center}

\begin{center}
J. E. Lawler, E. A. Den Hartog
\end{center}

\begin{center}
Department of Physics, University of Wisconsin, Madison, WI 53706;
\end{center}

\begin{center}
eadenhar@wisc.edu, jelawler@wisc.edu
\end{center}

\begin{center}
C. Sneden
\end{center}

\begin{center}
Department of Astronomy and McDonald Observatory, University of Texas, 
Austin, TX 78712; chris@verdi.as.utexas.edu
\end{center}

\begin{center}
and J. J. Cowan
\end{center}

\begin{center}
Homer L. Dodge Department of Physics and Astronomy, 
University of Oklahoma, Norman, OK 
73019; cowan@nhn.ou.edu
\end{center}

\begin{center}
ABSTRACT
\end{center}
Radiative lifetimes, accurate to $\pm $5{\%}, have been measured for 212 
odd-parity levels of Sm \textsc{ii} using laser-induced fluorescence. The 
lifetimes are combined with branching fractions measured using 
Fourier-transform spectrometry to determine transition probabilities for 
more than 900 lines of Sm \textsc{ii}. This work is the largest-scale 
laboratory study to date of Sm \textsc{ii} transition probabilities using 
modern methods. This improved data set has been used to determine a new 
solar photospheric Sm abundance, log $\varepsilon $ = 1.00 $\pm $ 0.03, from 
26 lines. The spectra of three very metal-poor, neutron-capture-rich stars 
also have been analyzed, employing between 55 and 72 Sm II lines per star. 
The abundance ratios of Sm relative to other rare earth elements in these 
stars are in agreement, and are consistent with ratios expected from rapid 
neutron-capture nucleosynthesis (the $r-$process).

Subject headings: atomic data --- stars: abundances stars: Population II --- 
Sun: abundances

\newpage 
\begin{center}
1. INTRODUCTION
\end{center}

Improved transition probability data for numerous Rare Earth (RE) spectra 
have been determined over the last decade. A complete review by Bi\'{e}mont 
{\&} Quinet (2003) has over 500 citations. Work on the second spectra 
(singly ionized) during the last decade includes: La~\textsc{ii} (Li {\&} 
Zhankui 1999, Zhiguo et al. 1999, Lawler et al. 2001a, Derkatch et al. 
2002), Ce \textsc{ii} (Palmeri et al. 2000, Zhang et al. 2001b), Pr 
\textsc{ii} (Ivarsson et al. 2001), Nd \textsc{ii} (Den Hartog et al. 2003), 
Sm \textsc{ii} (Scholl et al. 2002, Xu et al. 2003b), Eu \textsc{ii }(Zhang 
et al. 2000, Lawler et al. 2001c; Rostohar et al. 2001, Den Hartog et al. 
2002), Gd \textsc{ii} (Zhang et al. 2001a, Xu et al. 2003a), Tb~\textsc{ii} 
(Den~Hartog et al. 2001;  Lawler et al. 2001b), Dy \textsc{ii} (Curry et al. 
1997; Wickliffe et al. 2000), Ho \textsc{ii} (Den Hartog et al. 1999; Lawler 
et al. 2004), Tm \textsc{ii} (Anderson et al. 1996; Wickliffe {\&} Lawler 
1997, Rieger et al. 1999, Quinet et al. 1999a), Yb \textsc{ii} (Zhao et al. 
1996, Pinnington et al. 1997, Taylor et al. 1997, Bi\'{e}mont et al. 1998, 
Li et al. 1999, Yu {\&} Maleki 2000), Lu\textsc{ ii} (Den~Hartog, et al. 
1998; Quinet et al. 1999b; Fedchak et al. 2000). Many of the recent 
experimental studies have combined radiative lifetimes from laser induced 
fluorescence (LIF) measurements with emission branching fractions measured 
using a Fourier transform spectrometer (FTS). This approach to determining 
atomic transition probabilities in complex spectra has proved to be quite 
reliable.

A major motivation for the intense, multi-year effort on RE and other heavy 
element spectroscopy arises from stellar elemental abundance studies. 
High-resolution, high signal-to-noise spectra on a variety of targets from 
very large ground-based telescopes and the Hubble Space Telescope are 
providing data that are reshaping our views on the chemical evolution of our 
Galaxy. Old metal-poor Galactic halo stars provide a fossil record of the 
chemical make-up of our Galaxy when it, and the Universe, were very young 
(e.g., Gratton {\&} Sneden 1994; McWilliam et al. 1995; Cowan et al. 1995; 
Sneden et al. 1996, Ryan et al. 1996, Cayrel et al. 2004). Recent abundance 
determinations of heavy neutron capture ($n$-capture) elements in very 
metal-poor stars have yielded new insights on the roles of the $r$(apid)- and 
$s$(low)-processes in the initial burst of Galactic nucleosynthesis. 
Substantial progress is occurring due to improvements in both observational 
data and laboratory data needed for analysis of the spectra (e.g. Cowan et 
al. 2002; Sneden et al. 2003; Simmerer et al. 2004).

The RE's are among the most spectroscopically accessible of the $n$-capture 
elements. Large numbers of transitions of singly ionized RE species appear 
in the spectrum of the Sun and in stars over a significant temperature 
range. In addition to work on metal-poor stars, the solar abundances of 
several of these elements have been brought into agreement with meteoric 
abundances through improvements in the transition probability database (e.g. 
Bord et al. 1998; Den Hartog et al. 1998; Lawler et al. 2001b).

In this paper we apply LIF/FTS experimental techniques to determine accurate 
data for a very large number of ionized samarium transitions. In the recent 
study of Xu et al.(2003b)\textsc{, }theoretical branching fractions of Sm II 
have been combined with experimental lifetimes. The spectrum of Sm 
\textsc{ii} is very complex; there is a substantial breakdown of Russell- 
Saunders or {\it LS} coupling for most odd-parity levels in combination with 
extensive configuration mixing and relativistic effects. Therefore another 
one of the motivations for the work described herein is to test the 
relativistic Hartree Fock calculations by Xu et al. (2003b).

The above-mentioned studies on La~\textsc{ii}, Eu~\textsc{ii,} 
Tb~\textsc{ii}, and Ho \textsc{ii} (whose nuclei have odd atomic numbers Z 
and only odd mass numbers N in their naturally occurring isotopes) included 
measurements of both transition probabilities along with isotope shift 
and/or hyperfine structure data. Although Sm (Z = 62) has seven abundant 
isotopes, all but two of them have even N's. Therefore most of the blue and 
near UV lines of Sm \textsc{ii} are rather narrow with no resolved structure 
in our spectra with resolving powers up to 10$^{6}$. Several of the strong 
blue and near UV lines of Sm \textsc{ii} do have detectable structure and 
this opens the possibility of isotopic abundance determinations. Masterman 
et al. (2003) reported an extensive study of the isotopic and hyperfine 
structure of strong Sm \textsc{ii} lines in the blue region. In the present 
study, we have concentrated on determining transition probabilities. We 
measured radiative lifetimes using time-resolved laser-induced fluorescence 
for 212 odd-parity levels of Sm \textsc{ii}. These were combined with 
branching fractions measured using FTS data to yield {\it gf}-values for over 900 
transitions of Sm \textsc{ii}. This new data set was used to re-assess Sm 
abundances in the solar photosphere and in three $r-$process rich, metal-poor 
stars. The possibility of determining Sm isotopic abundances is discussed. 
We conclude with a brief review of the $n-$capture elemental abundances in 
metal-poor halo stars.

\begin{center}
2. OVERVIEW OF THE SECOND SPECTRUM OF SAMARIUM
\end{center}

Figure 1 is a partial Grotrian diagram for singly ionized Sm constructed 
from energy levels tabulated by Martin et al. (1978). The even-parity levels 
of the 4f$^{6}(^{7}$F)6s sub-configuration are all known. The 
4f$^{6}(^{7}$F)5d sub-configuration also appears to be complete in Martin 
et al., but a detailed analysis by Wyart {\&} Bauche-Arnoult (1981) led them 
to reject two ``high J'' levels listed by Martin et al. (Our figure labels 
this sub-configuration as $\sim $ known.) The first unknown even-parity 
levels are part of the 4f$^{6}(^{5}$D)6s and 4f$^{6}(^{5}$D)5d 
sub-configurations. These unknown even-parity levels have low J values and 
start about 15,500 cm$^{-1}$ according to Cowan Code (Cowan 1981) 
calculations (private communication: J.-F. Wyart 2005, D. Bord 2005). 
Fortunately, knowledge of even-parity levels below 20,000 cm$^{-1}$ is 
nearly complete. These low even-parity levels are sufficiently well isolated 
that they are relatively pure {\it LS} levels that can be assigned with confidence. 

The situation for odd-parity levels is not as satisfactory. The lowest 
odd-parity level is the 4f$^{7}$ $^{8}$S$_{7/2}$. This level is the only 
odd-parity level that is sufficiently isolated to be cleanly assigned. 
Martin et al. (1978) lists only tentative assignments for all higher 
odd-parity levels. Four extensively mixed configurations contribute to the 
band of odd-parity levels which starts at 21,250 cm$^{-1}$. These 
configurations are 4f$^{6}$6p, 4f$^{5}$6s5d, 4f$^{5}$6s$^{2}$, and 
4f$^{5}$5d$^{2}$. Xu et al. (2003b) point out that these four overlapping 
configurations have 13,628 levels and that the complete 4f$^{7}$ 
configuration, which also overlaps the other four configurations, has an 
additional 327 levels. A complete analysis of the low odd-parity 
configurations is a formidable task. It is not necessary to diagonalize a 
13,955 x 13,955 or larger matrix, because J (in addition to parity) is a 
good quantum number. Many of the levels in the five configurations are quite 
high in energy and do not strongly interact with the low odd-parity levels. 
However, even if the analysis is limited to the lowest sub-configurations 
built on the $^{7}$F and $^{6}$H parents (the 4f$^{6}(^{7}$F)6p, 
4f$^{5}(^{6}$H)6s5d, 4f$^{5}(^{6}$H)6s$^{2}$, and 
4f$^{5}(^{6}$H)5d$^{2}$ sub-configuration as denoted in Figure 1), this 
still includes more than 400 levels. The lack of assignment of the 
odd-parity levels is a problem which is discussed in more detail in {\S} 4.

All of the transitions studied in this work are from odd-parity upper 
levels. These levels decay primarily in the blue and near UV to the 
4f$^{6}(^{7}$F)6s even-parity levels. A substantial number of weaker 
transitions in the yellow and red to the 4f$^{6}(^{7}$F)5d even-parity 
levels were also measured in this work. The odd-parity levels in this study 
thus have significant 4f$^{6}(^{7}$F)6p or 4f$^{5}$6s5d components. 

\begin{center}
3. RADIATIVE LIFETIME MEASUREMENTS
\end{center}

Radiative lifetimes of 212 odd-parity levels of Sm \textsc{ii} have been 
measured using time-resolved laser-induced fluorescence (LIF) on a slow 
($\sim $5$\times $10$^{4}$~cm/s) atom/ion beam. The apparatus and technique 
are the same as used for many other species and have been described in 
detail elsewhere. Only a brief discussion is given here. The reader is 
referred to recent work in Eu \textsc{i}, \textsc{ii}, and \textsc{iii} (Den 
Hartog et al. 2002) for a more detailed description.

The beam of Sm atoms and ions is produced using a hollow cathode discharge 
sputter source. A large-bore hollow cathode is lined with samarium foil. A 
pulsed argon discharge, operating at $\sim $0.4~torr with 10~$\mu $s 
duration, 10~A pulses, is used to sputter the samarium. The hollow cathode 
is closed on one end except for a 1~mm hole, through which the samarium 
atoms and ions are extracted into a low pressure (10$^{-4}$~torr) scattering 
chamber. This beam is intersected at right angles by a nitrogen laser-pumped 
dye laser beam 1~cm below the cathode bottom. The laser is tunable over the 
range 2050 - 7200~{\AA} with the use of frequency doubling crystals, is 
pulsed at $\sim $30~Hz repetition rate with a $\sim $3~ns pulse duration, 
and has a 0.2~cm$^{-1}$ bandwidth. The laser is used to selectively excite 
the level to be studied. Selective excitation eliminates the possibility of 
cascade radiation from higher-lying levels. 

Fluorescence is collected in a direction orthogonal to both the laser and 
atomic/ionic beams through a pair of fused-silica windows which form an f/1 
optical system. Optical filters, either broadband colored glass filters or 
narrowband multi-layer dielectric filters, are typically inserted between 
the two lenses to cut down on scattered laser light and to block cascade 
radiation from lower levels. A RCA 1P28A photomultiplier tube (PMT) is used 
to detect the fluorescence. The signal from the PMT is recorded and averaged 
over 640 shots using a Tektronix SCD1000 digitizer. Data are recorded with 
the laser tuned on and off the excitation transition. A linear least-square 
fit to a single exponential is performed on the background-subtracted 
fluorescence decay to yield the lifetime of the level. The lifetime is 
measured twice for each level, using a different excitation transition 
whenever feasible. This redundancy helps ensure that the transitions are 
identified correctly in the experiment, classified correctly and are free 
from blends.

With only two exceptions, the lifetimes reported here have an uncertainty of 
$\pm $5{\%}. To achieve this level of fidelity and maintain it over the full 
dynamic range of the experiment (2~ns to 1.5~$\mu $s), the possible 
systematic errors in these measurements must be well understood and 
controlled. They include electronic bandwidth limitations, cascade 
fluorescence, Zeeman quantum beats and atomic motion time-of-flight effects, 
among others. These systematic effects are discussed in detail elsewhere, 
(See, for example, Den~Hartog et al. 1999; 2002) and will not be discussed 
further here. As a means of verifying that the measurements are within the 
stated uncertainties, we perform periodic end-to-end tests of the experiment 
by measuring a set of well known lifetimes. These cross-checks include 
lifetimes of Be \textsc{i} (Weiss 1995), Be \textsc{ii} (Yan et al. 1998) 
and Fe \textsc{ii} (Guo et al. 1992; Bi\'{e}mont et al. 1991), covering the 
range from 1.8--8.8 ns. An Ar~ \textsc{i} lifetime is measured at 27.85~ns 
(Volz {\&} Schmoranzer 1998). He \textsc{i} lifetimes are measured in the 
range 95 -- 220~ns (Kono {\&} Hattori 1984). 

The results of our lifetime measurements of 212 odd-parity levels of Sm 
\textsc{ii} are presented in Table 1. Energy levels are from the tabulation 
by Martin et al. (1978). Air wavelengths are calculated from the energy 
levels using the standard index of air (Edl\'{e}n 1953, 1966). The 
uncertainty of the lifetimes is $\pm $5{\%} with two exceptions which are 
noted in the table. 

Also presented in Table~1 is a comparison of our results with those from 
other LIF lifetime measurements available in the literature. We find that 
our lifetimes agree very well with the 35 lifetimes reported by Bi\'{e}mont 
et al. (1989) with the exception of 4 levels: 21508, 25940, 26821 and 28930 
cm$^{-1}$. In all four of these cases we measured our lifetime with 2 
different excitation transitions. In the case of the 21508~cm$^{-1}$ level, 
our lifetime is nearly an order of magnitude longer than what they measured. 
We speculate that they may have used the 4648.2 {\AA} transition to excite 
this level. We observe a very strong blend 0.08 {\AA} to the blue of this 
transition, which may have been inadvertently measured instead. No obvious 
explanation can be found for the factor of 2 - 2.5 disagreement on the 25940 
and 26821 cm$^{-1}$ levels. In addition to measuring each lifetime using two 
transitions, careful optical filtering was also used, so our confidence in 
our lifetimes is high. In the case of the 28930 cm$^{-1}$ level, they report 
a lifetime $\sim $25{\%} lower than ours. Here again, there is no obvious 
reason for the disagreement. With the three most serious discrepancies 
removed from the averages, we see a mean difference between our measurements 
and theirs of 1.0{\%} and an rms difference of 6.7{\%}. 

We find excellent agreement with the laser-fast beam measurements of 82 
lifetimes by Scholl et al. (2002). All our measurements are within 10{\%} of 
theirs, and the vast majority are well within 5{\%}. The mean difference 
between our measurements and theirs is 1.0{\%} and an rms difference of 
3.0{\%}. We do not see quite as good agreement with the LIF measurements of 
Xu et al. (2003b). Yet they are still all within 15{\%} of our measurements 
with a mean difference between our measurements and theirs of 5.2{\%} and an 
rms difference of 7.0{\%}. The agreement with the 18 laser-fast beam 
lifetimes of Vogel et al (1988) is excellent except for the 30{\%} 
discrepancy at 27285~cm$^{-1}$. When this discrepancy is omitted we see a 
mean difference between our measurements and theirs of 1.0{\%} and an rms 
difference of 1.8{\%}. They single out the 27285 cm$^{-1}$ level in their 
discussion as having comparatively large uncertainties. Our lifetime of this 
level was measured on two different transitions with narrowband, off-line 
filtering in both cases. Our confidence in this lifetime is high, especially 
as our value is in good agreement with those of both Bi\'{e}mont et al. 
(1989) and Scholl et al. (2002).

\begin{center}
4. BRANCHING FRACTIONS AND ATOMIC TRANSITION PROBABILITIES
\end{center}

The 1.0 meter FTS at the National Solar Observatory (NSO) was used in this 
work on Sm \textsc{ii}. This instrument is uniquely suited for 
spectroradiometry on complex RE atoms and ions. It provides: (1) a limit of 
resolution as small as 0.01 cm$^{-1}$, (2) wave number accuracy to 1 part in 
10$^{8}$, (3) broad spectral coverage from the UV to IR, and (4) the 
capability of recording a million point spectrum in 10 minutes (Brault 
1976). An FTS is insensitive to any small drift in source intensity since an 
interferogram is a simultaneous measurement of all spectral lines. The 
combination of branching fractions from FTS spectra with radiative lifetimes 
from LIF measurements has resulted in greatly improved atomic transition 
probabilities for the first and second spectra of many elements.

The emission sources for Sm spectra were commercially manufactured, sealed 
hollow cathode discharge (HCD) lamps with fused silica windows containing 
either argon or neon fills. We operated these lamps at currents 
significantly above the manufacturers' recommendation, but used forced air 
cooling to prevent overheating. The NSO 1.0 m FTS fitted with the UV beam 
splitter was used to record spectra during our February 2000 observing run. 
Spectra of the Ar{\-}filled lamp operating with a discharge current of 27 mA 
(61 co-adds), 27 mA (50 co-adds), 22 mA (4 co-adds), and 18 mA (68 co-adds), 
were taken with the ``super blue'' silicon diode detectors and no additional 
filtering. The term ``co-add'' refers to a coherently added interferogram. 
These spectra cover the 8,000 cm$^{-1}$ to 35,000 cm$^{-1}$ region with a 
limit of resolution of 0.053 cm$^{-1}$. Spectra of the Ne-filled lamp 
operating at 23 mA (10 co-adds) and at 17 mA (10 co-adds) were taken using 
an identical setup. An additional spectrum of the Ar-Sm lamp at 27 mA (50 
co-adds) was taken with the same set-up during our February 2002 observing 
run. Branching fraction measurements were made almost entirely on the Ar-Sm 
spectra: the Ne-Sm spectra were used only to separate Ar + Sm line blends. 
The three Ar-Sm spectra with the 50 or more co-adds and a lamp current of 27 
mA were most useful: the lower current Ar-Sm spectra were used primarily to 
check for optical depth errors on the very strongest lines. 

The HCD lamps used in this study operate with relatively low buffer gas 
pressures and thus are not in local thermodynamic equilibrium (LTE). This is 
not a problem because the absolute scale of a transition probability is 
provided by the radiative lifetime of the upper level in every case. Doppler 
broadening tends to dominate the emission line shapes in these low pressure 
lamps. Such narrow line shapes can produce radiation trapping or optical 
depth error. By comparing spectra from the Ar-Sm lamp operating at different 
currents, we verified that radiation trapping is not a problem. The 
sputtering rate of Sm in the hollow cathode discharge, and the total Sm 
density in the plasma, are strongly increasing function of discharge 
current.

The establishment of an accurate relative radiometric calibration or 
efficiency is critical to a branching fraction experiment. Detectors, 
spectrometer optics, lamp windows, and any other components in the light 
path or any reflections which contribute to the detected signal (such as due 
to light reflecting off the back of the hollow cathode), all have 
wavelength-dependent optical properties which must be taken into account 
when determining the ratio of line intensities at different wavelengths. 
Fortunately the radiometric efficiency of the FTS is a smoothly varying 
functions of wavelength. An excellent way to determine the relative 
radiometric efficiency of an FTS is to compare well-known branching ratios 
for sets of lines widely separated in wavelength, to the intensities 
measured for the same lines. Sets of Ar \textsc{i} and Ar \textsc{ii} lines 
have been established for this purpose in the range of 4300 to 35000 
cm$^{-1}$ by Adams {\&} Whaling (1981), Danzmann {\&} Kock (1982), 
Hashiguchi {\&} Hasikuni (1985), and Whaling et al. (1993), . These provide 
an excellent means of calibrating our FTS spectra since the argon lines are 
measured in the exact experimental arrangement and at the exact same time as 
are the Sm \textsc{ii} lines. A spectrum of a tungsten lamp, recorded before 
and after the 2002 Ar-Sm spectra, was used to interpolate between Ar 
reference lines to improve the relative radiometric calibration of the 2002 
data. The use of a tungsten lamp is of some value near the dip in the FTS 
sensitivity at 12,500 cm$^{-1}$ from the aluminum mirror coatings, and 
between 10,000 and 9,000 cm$^{-1}$ where the Si detector response is rapidly 
decreasing.

All possible transition wave numbers between known energy levels of Sm 
\textsc{ii} satisfying both the parity change and $\Delta $J = -1, 0, or 1 
selection rules were computed and used during analysis of FTS data. Energy 
levels from Martin et al. (1978) were used to determine possible transition 
wave numbers. Levels from Martin et al. (1978) are available in electronic 
form from Martin et al. (2000)\footnote{Available at 
http://physics.nist.gov/cgi{\-}bin/AtData/main{\_}asd}.

Branching fraction measurements were attempted on all 212 levels from the 
lifetime experiment, and were completed on 185 levels. Some of the levels 
for which branching fractions could not be completed had a strong branch 
beyond the UV limit of our spectra, or had a strong branch which was 
severely blended. Typically an upper level, depending on its J value, has 
about 30 possible transitions to known lower levels. More than 40,000 
possible spectral line observations were studied during the analysis of 7 
different Ar-Sm and Ne-Sm spectra. We set baselines and integration limits 
``interactively'' during analysis of the FTS spectra. A simple numerical 
integration routine was used to determine the un-calibrated intensities of 
Sm \textsc{ii} lines and selected Ar \textsc{ii} and Ar \textsc{i} lines 
used to establish a relative radiometric calibration of the spectra.

The procedure for determining branching fraction uncertainties was described 
in detail by Wickliffe et al. (2000). Branching fractions from a given upper 
level are defined to sum to unity, thus a dominant line from an upper level 
has small branching fraction uncertainty almost by definition. Branching 
fractions for weaker lines near the dominant line(s) tend to have 
uncertainties limited by their signal-to-noise ratios. Systematic 
uncertainties in the radiometric calibration are typically the most serious 
source of uncertainty for widely separated lines from a common upper level. 
We used a formula for estimating this systematic uncertainty that was 
presented and tested extensively by Wickliffe et al. (2000).

The problem of ``residual'' branches to currently unknown lower levels 
deserves some discussion because of the complexity of Sm \textsc{ii}. All of 
the upper levels in this study have strong branches in the blue and/or near 
UV to the 4f$^{6}(^{7}$F)6s sub-configuration. Observed weaker transitions 
in the yellow and/or red region to the 4f$^{6}(^{7}$F)5d 
sub-configuration, some of which are listed in Table 2, account for 0{\%} to 
44{\%} of the total decay. Typically, 15{\%} to 25{\%} of the total decay 
goes to the 4f$^{6}(^{7}$F)5d sub-configuration. We know that the unknown 
even-parity levels start about 15470 cm$^{-1}$. The lack of assignment for 
the odd-parity levels makes it difficult to estimate the strength of 
branches to the unknown levels. Two transition probabilities with similar 
dipole matrix elements scale in proportion to their frequencies cubed. This 
frequency scaling suppresses lower frequency transitions to the unknown 
even-parity levels. We estimate that branches to unknown lower levels from 
upper levels below 25,000 cm$^{-1}$ are negligible. A search of the near IR 
study of Sm \textsc{ii} by Blaise et al. (1969) supports this assessment. We 
measured in our study most of the strong near IR lines listed by Blaise et 
al. (1969). There is increasing risk of missing branches to unknown 
even-parity levels for higher odd-parity upper levels. Errors from missing 
branches are likely still covered by our total uncertainties for odd-parity 
upper levels in the 25,000 cm$^{-1}$ to 30,000 cm$^{-1}$ range. The 
situation for odd-parity upper levels in the 30,000 cm$^{-1}$ to 35,000 
cm$^{-1}$ range is less certain. Our transition probabilities from these 
upper levels could be too high by 10{\%} or perhaps somewhat more due to 
missing branches to unknown even-parity lower levels. Only one level above 
35,000 cm$^{-1}$ was included in our branching fraction study. The decay of 
this level at 38,505.66 cm$^{-1}$ is dominated by the near UV branch to the 
even-parity J= 8.5 level at 12,045.10 cm$^{-1}$. 

The difficulty in assigning the odd-parity levels is apparent in comments by 
Xu et al. (2003b). They report, ``{\ldots}according to our HFR calculations, 
the average purity, in {\it LS} coupling, of odd-parity levels below 23,000 
cm$^{-1}$ is found to be equal to 77{\%}, this value decreasing to 54{\%} 
for the levels situated between 23,000 cm$^{-1}$ and 25,000 cm$^{-1}$ and to 
32{\%} for those located between 25,000 cm$^{-1}$ and 35,000 cm$^{-1}$''. 
Our experimental results on the lifetimes and branching fractions from the 
odd-parity levels up to 35,000 cm$^{-1}$ should help in a more complete 
analysis and the eventual assignment of these levels. Cowan (1981) gave a 
lucid discussion of the difficulties in analysis and interpretation of RE 
spectra. He concludes that one should use a battery of experimental and 
theoretical aids.

Branching fractions from the FTS spectra were combined with the radiative 
lifetime measurements described in {\S}3 to determine absolute transition 
probabilities for 958 lines of Sm \textsc{ii} in Table 2. Transition 
probabilities for the weakest lines which were observed with poor 
signal-to-noise ratios are not included in Table 2, however these lines are 
included in the branching fraction normalization. Weaker lines are also more 
susceptible to blending problems. The effect of weaker lines becomes 
apparent if one sums all transition probabilities in Table 2 from a chosen 
upper level, and compares the sum to the inverse of the upper level lifetime 
from Table 1. Typically the sum of the Table 2 transition probabilities is 
75{\%} to 95{\%} of the inverse lifetime. Although there is significant 
fractional uncertainty in the branching fractions for these weaker lines, 
this does not have much effect on the uncertainty of the stronger lines 
which were kept in Table 2. Branching fraction uncertainties are combined in 
quadrature with lifetime uncertainties to determine the transition 
probability uncertainties in Table 2. This possible systematic error from 
missing branches to unknown even-parity lower levels is not included in the 
transition probability uncertainties listed in Table 2. Our estimates of 
such errors are very ``rough'' because of the lack of assignment of the 
odd-parity upper levels. We remind readers that any correction of errors 
from missing branches to unknown lower levels will always decrease tabulated 
transition probabilities. Such errors are: (1) thought to be negligible for 
upper levels below 25,000 cm$^{-1}$, (2) probably covered by our 
uncertainties for upper levels in the 25,000 cm$^{-1}$ to 30,000 cm$^{-1}$ 
range, and (3) may be 10{\%} or perhaps somewhat more for upper levels above 
30,000 cm$^{-1}$. Unknown branches do not affect the accuracy of the 
radiative lifetime measurements.

Although there have been a significant number of publications on Sm 
\textsc{ii} radiative lifetime measurements, we found only two which report 
original laboratory intensity measurements and either branching fractions or 
absolute transition probabilities. Relative intensity measurements by 
Meggers et al. (1961) were converted to absolute transition probabilities by 
Corliss {\&} Bozman (1962). Ward (1985) reported a formula for 
re-normalizing the Corliss {\&} Bozman (1962) transition probabilities. 
Cowley {\&} Corliss (1983) developed a formula for determining transition 
probabilities from line intensities published by Meggers et al. (1975) which 
are an updated version of the original Meggers et al. (1961) line 
intensities used by Corliss {\&} Bozman (1962). Saffman {\&} Whaling (1979) 
published a smaller, but high quality set of Sm \textsc{ii} branching 
fraction measurements made with a grating spectrometer and a photoelectric 
detection system. Most of the authors who report radiative lifetime 
measurements have generated transition probabilities by combining their 
lifetimes with branching fractions deduced from the Corliss {\&} Bozman 
(1962) transition probabilities or from measurements by Saffman {\&} Whaling 
(1979). The recent work by Xu et al. (2003b) is an important exception. They 
made a serious attempt to determine branching fractions from a relativistic 
Hartree Fock calculation. It is clear that a large scale effort to measure 
Sm \textsc{ii} branching fractions with an FTS is timely. 

Our first comparison is to the recent theoretical work by Xu et al. (2003b). 
Indeed, if the relativistic Hartree Fock method can be used with reliability 
to determine branching fractions in a spectrum as complex as Sm \textsc{ii}, 
then the need for further work using a FTS is much reduced. The next 
comparison is to branching ratios measured by Saffman and Whaling (1979). We 
omit a comparison to the oldest measurements by Meggers et al. (1961). There 
are already numerous published discussions of the problems in the Corliss 
{\&} Bozman (1962) transition probabilities (e.g. Obbarius {\&} Kock 1983). 

Figures 2, 3, and 4 show differences of log({\it gf}) values from Xu et al. (2003b) 
and log({\it gf}) values from our work, as functions of wavelength, transition 
probability, and upper transition energy, respectively . Xu et al. reported 
two sets of log({\it gf}) values. One set is based entirely on relativistic Hartree 
Fock calculations and the second set is normalized using experimental 
radiative lifetimes. The branching fractions of the second set are from the 
relativistic Hartree Fock calculations. Our comparison uses the second set 
because it is thought to be more accurate. The level of agreement in Figures 
2-4 is not as good as we hoped. The scatter in log({\it gf}) differences for the 
blue and near UV transitions is larger than the scatter for the yellow and 
red transitions (Figure 2). Difficulties in establishing an unambiguous 
correspondence between energy levels from the relativistic Hartree Fock 
calculations and experimental energy levels may be responsible for some of 
the scatter in the blue and near UV region. Only the log({\it gf}) differences 
plotted as a function of transition probability shown in Figure 3 shows a 
clear systematic trend. Our log({\it gf}) values for weaker lines tend to be smaller 
than the log({\it gf}) values from Xu et al. (2003b). The agreement ``on average'' 
in Figures 2 and 4 is from the use of experimental lifetimes to normalize 
both sets of transition probabilities. The use of slightly different 
radiative lifetimes from independent LIF measurements does not contribute 
much to the scatter as is shown below. The high density of odd-parity 
levels, the substantial breakdown of Russell-Saunders or {\it LS} coupling, and 
significant configuration mixing makes ab-initio calculations of Sm 
\textsc{ii} transition probabilities a formidable task.

Figures 5, 6, and 7 show similar comparisons of log({\it gf}) values from Saffman 
{\&} Whaling (1979) to log({\it gf}) values from our work. In order to make this a 
fair comparison we have normalized Saffman and Whaling's branching fractions 
using our experimental lifetimes. Saffman and Whaling did not have access to 
the large sets of radiative lifetimes from LIF measurements which are now 
available. The ordinates of Figures 2 through 7 are identical in order to 
facilitate comparisons. The experimental branching fractions from Saffman 
and Whaling are in better agreement with our branching fractions than are 
branching fractions from the relativistic Hartree Fock calculations. Average 
and root-mean-squared (rms) values of log({\it gf}$_{Xu})$ -- 
log({\it gf}$_{this\_expt})$ for the 84 lines in Figures 2 through 4 are --0.006 and 
0.41 respectively. If the comparison is limited to 11 lines listed in Table 
3 of Xu et al. (2003) that are common to all three investigations, then the 
average and rms differences are --0.17 and 0.40 respectively. Similarly, 
average and root-mean-squared values of log({\it gf}$_{SW})$ -- 
log({\it gf}$_{this\_expt})$ for the 48 lines in Figures 5 through 7 are --0.001 and 
0.17 respectively. Limiting the ``SW'' comparison to the 11 lines common to 
all three investigations yields average and rms differences of 0.005 and 
0.21 respectively. In addition to generally better agreement, no systematic 
trends with wavelength, transition probability, or upper transition energy 
can be discerned in the SW comparison of Figures 5 through 7. Differences 
between our branching fractions and those from Saffman {\&} Whaling can, in 
many cases, be traced to line blends which were not resolved with Saffman 
{\&} Whaling's grating spectrometer. Indeed, the high spectral resolving 
powers of the Kitt Peak 1.0 m FTS provided part of Prof. Whaling's 
motivation for pioneering its use in branching fraction measurements (Adams 
{\&} Whaling 1981).

Figure 8 and 9 are comparisons of various sets of radiative lifetimes from 
LIF measurements to our measurements. The ordinate in Figure 8 has a similar 
logarithmic scale as Figures 2-7. Figure 8 clearly demonstrates that the 
absolute scale established using LIF lifetimes is not contributing much of 
the log({\it gf}) scatter. Figure 9, with an expanded ordinate, reinforces our claim 
of 5{\%} total uncertainty on our lifetime measurements. Most of the points 
in Figure 9 have ordinate values between --0.02 (-5{\%}) and +0.02 (+5{\%}). 
We conclude that for complex spectra such as Sm II, branching fraction 
uncertainties are greater than the lifetime uncertainties except for the 
dominant branches.

\begin{center}
5. SOLAR AND STELLAR SAMARIUM ABUNDANCES
\end{center}

In this section we describe application of the new Sm\textsc{ ii} transition 
probability data to the solar spectrum and to the spectra of a few very 
metal-poor ([Fe/H] $<$ -2)\footnote{ We adopt standard stellar spectroscopic 
notations that for elements A and B,\par [A/B] = 
log$_{10}$(N$_{A}$/N$_{B})_{star}$ - log$_{10}$(N$_{A}$/N$_{B})_{sun}$, 
for abundances relative to solar ones, and \par log $\varepsilon $(A) = 
log$_{10}$(N$_{A}$/N$_{H})$ + 12.0, for absolute abundances. Overall 
metallicity is equated to the stellar [Fe/H] value.} stars. As in previous 
papers of this series, we chose for detailed investigation three metal-poor 
stars that are enriched in products of rapid $n-$capture ($r-$process) 
nucleosynthesis: HD 115444 ([Fe/H] = -2.9, [Eu/Fe] = +0.8, Westin et al. 
2000); BD+17$^{o}$3248 ([Fe/H] = -2.1, [Eu/Fe] = +0.9, Cowan et al. 2002), 
and CS 22892-052 ([Fe/H] = -3.1, [Eu/Fe] = +1.5, Sneden et al. 2003). Our Sm 
\textsc{ii} abundance analysis followed the methods used in previous papers 
of this series, most closely resembling those employed for Nd \textsc{ii }by 
Den Hartog et al. (2003).

\begin{center}
5.1 Line Selection
\end{center}

The new laboratory study yielded nearly a thousand potentially useful Sm 
\textsc{ii} lines. Selection of appropriate transitions for abundance 
analysis in the Sun and stars was the next task. We did not examine each of 
the Sm \textsc{ii} lines of Table 2 in each of the program stars; more 
efficient means to the same end were adopted. Initial inspection of the 
solar and stellar spectra revealed that Sm \textsc{ii} lines are usually 
very weak in all of our stars. For example consider 4424.35 {\AA}, one of 
the strongest ($\chi $ = 0.48 eV, log({\it gf}) = +0.14) relatively unblended lines 
of this species. Its measured equivalent width ({\it EW}) is 31 m{\AA} in CS 
22892-052, 25 m{\AA} in BD+17$^{o}$3248, 9 m{\AA} in HD 115444, and $<$15 
m{\AA} (very contaminated by other absorption features) in the Sun. This 
implies that the reduced widths ({\it RW = EW/$\lambda $}) for nearly all Sm \textsc{ii} lines are 
small: log({\it RW}) $<$ -5.1, placing these transitions on the linear part of the 
curve-of-growth. Therefore line saturation did not pose much difficulty in 
our abundance analysis. 

In general, the absorption strengths of weak lines that arise from a single 
species (e.g., Sm \textsc{ii}) vary directly with their transition 
probabilities modified by their Boltzmann factors. Thus for a given star, 
relative log({\it RW}) is proportional to log({\it gf}) --{\it  $\theta \chi $}, where {\it $\chi $} is in units of eV and 
inverse temperature {\it $\theta $ = }5040/T. Adopting T$_{eff}$ = 5778 K for the Sun (e.g. 
Cox 2000), 4650 K for HD 115444 (Westin et al. 2000), 5200 K for 
BD+17$^{o}$3248 (Cowan et al. 2002), and 4800 K for CS 22892-052 (Sneden et 
al. 2003) yields {\it $\theta $}$_{eff}$ = 0.87, 1.08, 0.97, and 1.05, respectively, or 
$<${\it  $\theta $} $>$ = 0.99 $\approx $ 1.0 for our stars. Using this mean inverse 
temperature value, we computed relative strength factors for all Sm 
\textsc{ii} lines. These are displayed in Figure 10.

For the strong 4424 {\AA} line, log({\it gf}) - {\it $\theta \chi $} = +0.14 - 0.48 = -0.34. We assume 
that this relative strength corresponds to the largest reduced width of the 
line in our program stars: log({\it RW}) = log(0.031/4424) = -5.15 in CS 22892-052. 
A reasonable lower limit for unambiguous detection of the 4424 {\AA} line in 
the Sun or stars, given data of high resolution and signal-to-noise, would 
be $\approx $1.5 m{\AA}, or log({\it RW}) $\approx $ -6.45. Then assuming that the 
4424 {\AA} line is unsaturated in all cases, the reduced width for detection 
implies a relative strength of -0.34 + 5.15 - 6.45 = -1.64, which should 
represent an approximate strength detection limit for Sm \textsc{ii} 
features. This limit is denoted by a horizontal dashed line in Figure 10.

Of the total list of 958 lines, 410 of them are stronger than the detection 
limit relative strength value. Concentrating mainly on CS 22892-052 and 
BD+17$^{o}$3248, which have larger $n$-capture elemental abundances than does 
HD 115444, we made many searches for lines somewhat below the strength 
detection limit in our spectra, but failed to recover any additional Sm 
\textsc{ii} lines for abundance analyses. Therefore we discarded lines with 
log({\it gf}) - {\it $\theta \chi $} $<$ -1.64, thus eliminating about 550 weaker transitions from 
further consideration. We caution the reader that our estimated Sm 
\textsc{ii} line strength detection limit applies only to the present 
program stars. A more favorable situation can easily be imagined: in the 
spectra of cool, relatively metal-rich giant stars all Sm \textsc{ii} lines 
will be much stronger, and thus the relative detection limit will be lower.

We examined the remaining 410 strongest lines. All of these transitions lie 
at wavelengths $\lambda   <$ 5000 {\AA}, where blending with lines of other 
atomic and molecular features must be treated with care. Indeed, about 240 
of these Sm \textsc{ii} lines occur below $\lambda  <$ 4000 {\AA}, where 
essentially no unblended line of interest can be detected in the solar 
spectrum. The majority of the 410 relatively strong lines were thus quickly 
discarded because they were totally masked or severely compromised by 
transitions of other species.

To give readers a better understanding of this process of elimination, we 
discuss here just the four Sm \textsc{ii} lines with the largest relative 
strength factors: 3568.27, 3592.60, 3609.49, and 3634.27 {\AA}, which are 
specially marked in Figure 10. All of these lines are identified as Sm II in 
the solar spectral atlas of Moore et al. (1966). The 3568 {\AA} line appears 
to be relatively clean. Moore et al. list the following transitions here: 
3568.14 {\AA} (Zr \textsc{ii}, {\it $\chi $} = 0.80 eV: {\it EW} = 6.5 m{\AA}), 3568.25 {\AA} (Sm 
\textsc{ii}, 0.48 eV: 28 m{\AA}), and 3568.31 {\AA} (unidentified: 12 
m{\AA}). However, not only is the {\it EW} of the Sm \textsc{ii} solar line large, 
but its measured wavelength is too blue to be attributed solely to the 
desired Sm \textsc{ii} transition. This line is blended in the Sun, and 
examination of the current version of Kurucz's (1998) atomic and molecular 
line compendium\footnote{ Available at http://kurucz.harvard.edu/} reveals 
only one plausible contaminant: 3568.24 {\AA} (Fe \textsc{i}, 2.48 eV). 
Lacking definitive identification, we assumed that Fe \textsc{i} is the 
correct blending agent, and retained this Sm \textsc{ii} feature for 
abundance analyses. The 3592 {\AA} line is a relatively minor part of a 
complex blend consisting mainly of 3592.48 {\AA} (Fe \textsc{i}, 2.59 eV: 42 
m{\AA}), 3592.60 {\AA} (Sm \textsc{ii}, 0.38 eV: 15 m{\AA}), 3592.68 {\AA} 
(Fe \textsc{i}, 3.24 eV: 79 m{\AA}), and 3592.90 {\AA} (Fe \textsc{i}, 2.20 
eV, and Y \textsc{i}, 0.00 eV: 48 m{\AA}). The contaminants are too large in 
solar/stellar spectra to use this line. The 3609 {\AA} transition lies in 
the wing of an extremely strong Fe \textsc{i} line (1.01 eV: 1046m{\AA}), 
and is buried in a local complex blend: 3609.33 {\AA} (Ni \textsc{i}, 0.11 
eV: 69 m{\AA}), 3609.47 {\AA} (Fe \textsc{i}, 2.86 eV, Cr \textsc{i}, 2.54 
eV, Sm \textsc{ii}, 0.28 eV: 42 m{\AA}), and 3609.56 {\AA} (Pd \textsc{i}, 
0.96 eV: 13 m{\AA}). It also had to be discarded. Finally, the 3634.27 {\AA} 
(0.18 eV: 7.5 m{\AA}) line is sandwiched between 3634.20 {\AA} 
(unidentified: 58 m{\AA}) and 3634.33 {\AA} (Fe \textsc{i}, 2.94 eV: 136 
m{\AA}), rendering it unusable. Therefore of the four strongest Sm 
\textsc{ii} lines, only 3568 {\AA} survived preliminary inspection to be 
employed in our abundance analyses. These line blending considerations 
eliminated all but $\sim $100 Sm \textsc{ii} lines. The remaining 
transitions were subjected to a more complete analysis.

\begin{center}
5.2 The Solar Photospheric Samarium Abundances
\end{center}

The vast majority of Sm \textsc{ii} transitions surviving the line selection 
process had residual blending and/or continuum placement concerns in the 
solar spectrum. Therefore we chose not to use equivalent width analyses for 
any lines. Instead we determined samarium abundances entirely from synthetic 
spectrum computations, in the same manner described in previous papers of 
this series.

Briefly, we assembled lists of atomic and molecular (CH and CN) lines in 4-6 
{\AA} surrounding each Sm \textsc{ii} transition, using the Kurucz (1998) 
line database, and supplementing with some identifications in the Moore et 
al. (1966) solar atlas. We then used the current version of the LTE line 
analysis code MOOG (Sneden 1973) to generate synthetic spectra, adopting the 
Holweger {\&} M\"{u}ller (1974) empirical solar model atmosphere for these 
computations. Standard solar abundances (Grevesse {\&} Sauval 1998, 2002; 
Lodders 2003) were assumed for most elements. Solar abundances of elements 
newly determined in this series (La, Nd, Eu, Tb, Ho, Pt) were taken from 
those papers listed in {\S} 1. Transition probabilities of these elements 
were adopted without change from the respective laboratory analyses, as were 
those of Ce \textsc{ii} (Palmeri et al. 2000) and of course Sm \textsc{ii} 
(this study).

We took the observed solar photospheric spectra from the center-of-disk 
intensity spectral atlas of Delbouille et al. (1973)\footnote{ Available at 
http://mesola.obspm.fr/solar{\_}spect.php}. Iterative comparisons of 
synthetic and observed center-of disk solar spectra yielded adjustments to 
the transition probabilities of contaminating spectral features. Molecular 
line strengths were altered as a group by varying abundances of the 
molecular constituent atoms. For absorptions present in the observed solar 
spectrum that have no plausible atomic or molecular identifications, we 
arbitrarily assumed that they were Fe \textsc{i} lines with excitation 
potentials {\it  $\chi $} = 3.5 eV and transition probabilities arbitrarily set to match 
the observed features. These trial synthetic spectrum computations served 
also to eliminate $\sim $20 more Sm \textsc{ii} lines as unsuitable for both 
solar and stellar abundance computations for one or more of the reasons 
discussed above.

Final synthetic/observed matches to the approximately 80 potentially useful 
Sm \textsc{ii} transitions yielded solar photospheric abundances for 36 of 
them. The lines not used in the solar analysis were retained for the stellar 
analysis described in the next section. The solar abundances for each line 
are listed in Table 3, and plotted as a function of wavelength in the top 
panel of Figure 11. From these 36 lines, we derived a mean abundance of 
log$\varepsilon $(Sm)$_{Sun}$ = +1.00 $\pm $ 0.01 ($\sigma $ = 0.05). 
Abundance uncertainties for the solar Sm \textsc{ii} lines are comprised of 
the effects of (internal) line profile fitting issues and (external) scale 
factors. Repeated matches of observed to synthetic spectra suggested that 
abundances for each could be estimated on average to $\pm $ 0.02 dex. For 
many photospheric Sm \textsc{ii} lines, contamination by other features adds 
to this line profile fitting uncertainty, again estimated on the basis of 
repeated trial syntheses to be $\pm $ 0.02 dex. Adding these probable errors 
in quadrature to the estimated $\pm $ 0.02 dex typical log({\it gf}) uncertainties 
for strong lines yields a total internal line-to-line scatter error of 
$\approx   \pm $ 0.04 dex, consistent with the observed $\sigma $ = 0.05. 

Overall scale errors can arise from other atomic data uncertainties and 
model atmosphere choices. Samarium, like all other rare earth elements, has 
a relatively small first ionization potential, 5.644 eV (e.g., Grigoriev 
{\&} Melikhov 1997). Therefore it exists almost entirely as Sm \textsc{ii} 
in the solar photosphere (and in the line-forming atmospheric layers of our 
metal-poor program giants), so that Saha-fraction corrections are 
negligible. Thus the derived Sm abundance varies almost linearly with the Sm 
\textsc{ii} partition function (which enters through the Boltzmann 
equation). We employed the most recent atomic energy level data (Martin et 
al. 1978, Martin et al. 2000) to computed partition functions for Sm 
\textsc{ii}, finding no change $>$ 0.01 dex from the temperature-dependent 
polynomial representations of Irwin (1981). There are, however, many 
experimentally unknown levels of both parities above 20,000 cm$^{-1}$ as 
discussed in {\S} 2. The existence of these levels is certain from counting 
angular momentum projections. The best approach to correcting a partition 
function for unknown levels is to perform a Cowan Code (Cowan 1981) 
calculation and to merge (eliminating duplicates) the resulting theoretical 
energy level list with the experimental list (eg. Bord {\&} Cowley 2002). 
Prof. Don Bord (private communication 2005) has done such a calculation for 
Sm \textsc{ii} and shared selected results with us. The Cowan code approach 
yields a partition function 2.2{\%} (5.6{\%}) larger than that from the 
Martin et al. energy levels at 5000K (6000K). Although we have not included 
this ``unknown level'' correction to the partition function in our abundance 
studies, its effect is to increase Sm abundance for the Sun by 0.01 dex. The 
correction is similar for the warmest of the r-process rich stars, 
BD+17$^{o}$3248, discussed in {\S} 5.3.

To assess the influence of solar model atmosphere choice, we repeated the 
abundance computation for the 4424 {\AA} line with Kurucz (1998) and 
Grevesse {\&} Sauval (1999) models, finding abundance differences of -0.02 
and -0.03 dex with respect to those derived with the Holweger {\&} 
M\"{u}ller (1974) model. These differences are nearly identical to those 
determined for other rare earth ions in the previous papers of this series. 
Therefore abundance scale errors appear to be very small, of order 0.03 dex. 
Considering internal line-to-line scatter uncertainties (negligible in the 
mean value, given the large number of lines used) and external scale 
uncertainties, we suggest that log$\varepsilon $(Sm)$_{Sun}$ = +1.00 $\pm $ 
0.03 be adopted as the solar photospheric samarium abundance.

In recent compilations of solar-system data, Grevesse {\&} Sauval (1998) 
have recommended a meteoritic abundance of log$\varepsilon $(Sm)$_{Sun}$ = 
+0.98 $\pm $ 0.02, while Lodders (2003) has recommended +0.95 $\pm $ 0.04. 
Our new photospheric abundance is consistent with these meteoritic values to 
within the stated uncertainties. 

Several previous studies over the last 30 years have discussed the solar 
photospheric Sm abundance. Line-by-line abundances from these papers are 
displayed in the lower panel of Figure 11. Most of the previous analyses 
employed very few Sm \textsc{ii} transitions: Andersen et al. (1975), log 
$\varepsilon $(Sm)$_{Sun}$ = +0.72 from one line; Saffman {\&} Whaling 
(1979), +0.80 $\pm $ 0.11 from four lines out of six attempted; Vogel et al. 
(1988), +1.02 $\pm $ 0.11 from one line out of four; and Youssef {\&} Khalil 
(1989), +0.99 $\pm $ 0.05 from two lines out of three. They will not be 
considered further here. As discussed in previous sections, Bi\'{e}mont et 
al. (1989) combined their own lifetime measurements with branching fractions 
taken mainly from Corliss {\&} Bozman (1962) to determine transition 
probabilities for nearly 40 Sm \textsc{ii} lines, and derived a photospheric 
abundance of 

log $\varepsilon $(Sm)$_{Sun}$ = +1.00 $\pm $ 0.03 ($\sigma $ = 0.14) from 
an {\it EW} analysis of 26 of these lines. Our analysis is clearly in excellent 
agreement with the Bi\'{e}mont et al. study. The factor of three decrease 
between the line-to-line scatter of their study and the present one appears 
to arise from use of improved {\it gf} values and adoption of a synthetic-spectrum 
approach in the solar Sm analysis .

\begin{center}
5.3 Samarium Abundances in $r-$Process-Rich Stars
\end{center}

We next analyzed Sm \textsc{ii} lines in HD 115444, BD+17$^{o}$3248, and CS 
22892-052. We computed Sm abundances via synthetic spectrum computations 
with the line lists developed for the solar analysis. The combination of 
overall metal deficiency and relative $n$-capture-element enhancement in these 
stars produces many relatively unblended Sm \textsc{ii} lines, allowing 
reliable abundances to be determined for many transitions that are 
hopelessly masked in the solar spectrum. In Table 3 we list the abundances 
from individual lines in the three stars, and in Table 4 we summarize 
previously-published and new mean Sm abundances. The line-to-line scatters 
are all small: $\sigma $ = 0.05 - 0.07, decreased from the typical 0.15 
values of the original studies. In Figure 12 we demonstrate this result for 
BD+17$^{o}$3248, as was done for Nd \textsc{ii} lines in Figure 1 of Den 
Hartog et al. (2003). Repeating also the numerical experiments of that 
paper, we have computed line-by-line Sm abundance differences between the 
metal-poor giants and the Sun, between HD 115444 and BD+17$^{o}$3248, and 
between CS 22892-052 and BD+17$^{o}$3248. In all cases we found $\sigma $ 
{\{}$\Delta $ log $\varepsilon $(Sm) {\}} $\ge $ 0.07. That is, no reduction 
in $\sigma $ values was achieved by comparing abundances between stars on a 
line-by-line basis. This indicates that the abundance scatters are dominated 
by various factors in the stellar analyses; uncertainties in transition 
probabilities appear to be negligible contributors here. 

\begin{center}
6. ABUNDANCES OF $n$-CAPTURE ELEMENTS IN METAL-POOR HALO STARS
\end{center}

One of the main motivations of the new laboratory experiments is to provide 
increasingly more accurate determinations of the heavy $n-$capture element 
abundances in the metal-poor halo stars. Understanding these stellar 
distributions, and in particular comparing them to the (total and 
$r$-process only) solar system abundance distributions, also provides fresh new 
insights into the astrophysical conditions, the nuclear processes and the 
stellar sites for heavy element nucleosynthesis. Since the halo stars are so 
old and were formed so early, such abundance comparisons and analyses may 
also help to identify the earliest Galactic stellar generations. 

In Figure 13 we summarize the $n-$capture abundances in the atomic number range 
56 $\le $ Z $\le $ 68 for the solar system and for the three very metal-poor 
([Fe/H] $<$ -2) halo giant stars CS 22892-052, HD 115444 and 
BD+17$^{o}$3248. We have plotted the abundance differences (log observed 
abundance -- log solar system $r$-process only value) for each element in each 
star. For this comparison we have normalized the abundance distributions of 
all three stars at the $r-$process element Eu. The solar system elemental 
$r$-process abundance distribution was obtained by summing the individual 
$r$-process isotopic abundance contributions, based upon the so-called standard 
model (see Simmerer et al. 2004 and discussion below in {\S}7). Perfect 
agreement between the stellar and solar $r$-process values would result in a 
difference of zero, and thus would fall on the solid horizontal line in the 
figure. We also compare the abundance differences for Sm and the other 
$n$-capture elements in the top panel of the figure with the total Solar System 
meteoritic abundance values (dotted-line curve) recommended by Lodders 
(2003). The stellar abundances in this top panel are those of the original 
papers on these stars by our group (Westin et al. 2000, Cowan et al. 2002; 
Sneden et al. 2003). A large scatter, reflected in the large error bars, is 
seen especially for elements Nd, Sm, and Ho in the stellar data. 

The bottom panel of Figure 13 reflects the recent efforts to obtain improved 
laboratory data for various elements of astrophysical interest. In this 
panel we have added the most recent solar photospheric abundances (black 
dots) in order to compare them with meteoritic values. The new laboratory 
transition probabilities for Nd (Den Hartog et al. 2003) and Ho (Lawler et 
al. 2004) have resulted in excellent agreement seen in this panel for the 
abundances (with respect to the solar $r$-process values) for those elements in 
CS 22892-052, HD 115444 and BD+17$^{o}$3248.\footnote{Beyond the RE group, 
thus not shown in Figure 13, is the very heavy r-process-dominated element 
Pt, which is accessible to high resolution UV spectroscopy. A new laboratory 
analysis of Pt (Den Hartog et al. 2005) has resulted in a similar 
improvement of the abundance of this element in BD+17$^{o}$3248.} For Sm the 
comparison between the top and bottom panels of Figure 13 indicates 
graphically the new concordance of the abundance data for this element - 
with very small uncertainties the differences between the stellar and solar 
$r-$values are identical in CS 22892-052, HD 115444 and BD+17$^{o}$3248. It is 
clear now that the abundance ratio of Sm/Eu is the same for all three stars 
(see also Table 4). This strongly indicates that both Sm and Eu were 
synthesized in the $r$-process {\it only}, and with relative solar system values, in 
nucleosynthesis sites that ejected these elements into the interstellar 
medium early in the history of the Galaxy and are seen now in all three of 
these old halo stars. This is in contrast to Sm in solar system material, 
which has an $r-$process fraction of 67{\%}. (Eu is always made almost entirely 
in the $r-$process.)

The data of Figure 13 demonstrate that the differences in the abundances for 
the elements below Sm (Z $<$ 62) are not consistent with the total (dotted 
line curve) photospheric solar abundances (Lodders 2003), but instead fall 
mostly on the horizontal solid line, indicating an $r-$process only origin. This 
follows since in solar system material, elements such as Ba and La are 
predominantly synthesized in the $s-$process, not the $r-$process. For elements 
above Sm (Z $>$ 62), the total solar system abundances are overwhelmingly 
from $r-$process nucleosynthesis - Eu, Tb and Ho are 97{\%}, 93{\%}, 94{\%} 
$r-$process, respectively (Simmerer et al. 2004) - so not surprisingly the solar 
and $r-$process curves are almost coincident. The agreement between the 
abundance data of the heavier $n-$capture elements and both of these curves is 
again an indication of an $r-$process only synthesis history, and a further 
indication that the stellar and relative solar system $r-$process abundances are 
consistent. Previous abundance determinations of Sm in the three metal-poor 
halo stars (as shown in the upper panel of Figure 13 ) had large errors bars 
and were scattered from the $r-$process only up to and on the total solar system 
curve. With the new more accurate atomic physics data the Sm (relative to 
solar system $r-$process only) abundances for all three stars now clearly lie 
below the total solar system abundances and are consistent with the 
$r-$process only solar curve. 

This agreement between the $n-$capture element abundances in many metal-poor 
halo stars and the solar system $r-$process abundance distribution has been 
noted for some time (see the reviews of Truran et al. 2002; Sneden {\&} 
Cowan 2003; Cowan {\&} Thielemann 2004). The $s-$process nuclei and elements 
that contribute to solar system matter are produced in low- and 
intermediate-mass stars that evolve over very long time periods (Busso et 
al. 1999). Thus, early in the history of the Galaxy, when the most 
metal-poor halo stars formed, the $s-$process could not have been responsible 
for major element formation. Therefore, the predominant early Galactic 
synthesis must have resulted from the $r-$process, and presumably from rapidly 
evolving sites such as core-collapse supernovae (Cowan {\&} Thielemann 
2004).

While it has been clear that there is a general consistency between the 
$n-$capture elements in the metal-poor halo stars and the relative (or scaled) 
solar system $r-$process abundance distribution, this agreement has for some 
elements, and in some cases, been only approximate with fairly large 
abundance uncertainties. The new experimental atomic transition data for 
individual elements - in this case Sm - has now made that agreement much 
more precise. In fact the scatter has become so low and the agreement has 
become so good that the abundance data might now possibly be employed to 
constrain predictions for the solar system abundances. In the particular 
case of Sm it is seen in Figure 13 that the abundances of all three halo 
stars lie above the solar system $r-$process prediction - here the standard 
model value. This might suggest that the value for the solar system 
$r-$process (and likewise the $s-$process) contribution to Sm should be reassessed. 

Finally, we note that there are a few exceptions to the excellent agreement 
for element-to-element abundances in these stars and with the solar system 
$r-$process abundances. Gd for example (in Figure 13 ) shows considerable 
abundance scatter suggesting a need for new experimental atomic data for 
that element. Likewise, there are several other cases where improved atomic 
laboratory data would make more precise the abundance determinations of 
certain $n-$capture elements in the metal-poor halo stars. 

\begin{center}
7. ISOTOPIC AND HYPERFINE SUBSTRUCTURE CONSIDERATIONS FOR SM \textsc{II}
\end{center}

Samarium has seven abundant naturally occurring isotopes: $^{144}$Sm 
(3.1{\%} of the solar-system elemental abundance), $^{147}$Sm (15.0{\%}), 
$^{148}$Sm (11.2{\%}), $^{149}$Sm (13.8{\%}), $^{150}$Sm (7.4{\%}), 
$^{152}$Sm (26.8{\%}), and $^{154}$Sm (22.7{\%}). Isotopes 147 and 149 can 
have complex hyperfine structures, and transition substructures must exist 
for all Sm \textsc{ii} lines to some degree. However, for lines used in the 
present abundance analyses, the combined hyperfine and isotopic 
substructures are reasonably compact. On the FTS laboratory spectra, 
measured full-width-at-half-maxima of completely unresolved Sm \textsc{ii} 
lines are $\approx $ 0.06 cm$^{-1}$, consistent with the 0.053 cm$^{-1}$ 
spectral resolution limit given in {\S}4. In wavelength units this 
corresponds to $\Delta \lambda   \approx $ 0.01 {\AA} for lines near 4000 
{\AA}. As indicated by the ``Comment'' column in Table 3, most transitions 
used in our abundance analyses show no intrinsic broadening in excess of 
this value in our lab spectra. The Doppler smearing of lines in the stellar 
atmospheres of this study is $\surd $(2k$_{B}$T/M +v$_{t}^{2})  \ge $ 2 
km s$^{-1}$ or $\Delta   \lambda  \ge $ 0.03 {\AA}, much larger than the 
intrinsic line broadening. Therefore for present purposes most Sm 
\textsc{ii} lines can be treated as single transitions. Those lines with 
detectable intrinsic broadening in excess of 0.01 {\AA} are noted with 
``hfs'' in the ``Comment'' column.

The relatively strong 4424 {\AA} line discussed earlier is one of the Sm 
\textsc{ii} transitions that exhibits significant hyperfine/isotopic 
splitting on our lab spectra. Three obvious components of this line are 
blended together, with a total wavelength spread of $\approx $ 0.04 {\AA}. 
Comprising this blend are 21 components each of isotopes 147 and 149, and 
single components of the remaining five isotopes. Using the isotopic and 
hyperfine structure data of Masterman et al. (2003) we synthesized the 4424 
{\AA} line in all four stellar spectra. As expected given the previous 
discussion about the intrinsic weakness of Sm \textsc{ii} lines in our 
stars, we found little change to abundances determined with the single-line 
approximation to this feature. Since the 4424 {\AA} line is one of the 
strongest of all Sm \textsc{ii }transitions, further consideration of 
hyperfine/isotopic substructure seemed unwarranted here. More careful 
treatment would be required for accurate elemental abundance determinations 
in stars with deeper, more saturated Sm \textsc{ii} lines.

Although these tests with the 4424 {\AA} line indicate that it will be quite 
difficult to determine Sm isotopic abundances, the possibility still merits 
some discussion. Many 5d -- 6p yellow-red lines of Sm \textsc{ii} have 
appreciably wider structure than the 6s --6p blue-UV lines. Continued 
improvements in telescopes and spectrometers or the discovery of a more 
favorable star might make it possible to determine a Sm isotopic abundance, 
and so we include here a summary of the $s-$ and $r-$process contributions to Sm and 
some predictions of possibly observable effects.

We neglect the small (3{\%} of the solar system) contribution to Sm from the 
$p-$process isotope $^{144}$Sm in the following discussion. We show in Table 5 
the individual breakdown of the contribution by process. The $s-$process 
additions to the solar system abundances were first determined based upon 
one of several approaches. The so-called standard or ``classical'' 
(empirical) model assumes a smooth fit (smooth behavior) of the ``$\sigma 
_{n}$ N$_{s}$'' curve ($i.e.$ the product of the $n-$capture cross-section $\sigma 
_{n}$ and $s-$process abundances N$_{s})$ to the solar system $s-$process 
abundances. Determinations of the neutron-capture cross sections (see e.g. 
K\"{a}ppeler et al. 1989) then allows the direct determination of the 
N$_{s}$ contributions to each isotope (see Burris et al. 2000; Simmerer et 
al. 2004). An alternative approach employs more detailed (low-mass AGB) 
stellar model and nucleosynthesis calculations to obtain the isotopic 
$s-$process solar system abundances (see Arlandini et al. 1999). Subtracting 
these $s-$process isotopic abundances from the total solar abundances determines 
the individual $r-$process contributions, or residuals. We note that 
experimental determinations of individual $r-$process abundance contributions 
are, in general, not possible at this time.

We have listed the values from both the standard and stellar models in Table 
5. The abundances for the $s-$ and $r-$process contributions are based upon the Si = 
10$^{6}$ scale. We have also listed the percentage contribution by 
individual isotope to the total elemental $s-$ and $r-$process abundances (i.e., the 
vertical columns add up to 100{\%} in those particular columns). It is clear 
from the table that both the standard and stellar models give very similar 
isotopic abundance predictions for the $s$- and$ r-$process mixtures for Sm.

It may be possible to observe the isotopic mixture of Sm in a metal-poor 
halo star. Lambert {\&} Allende Prieto (2002) observed the isotopic mixture 
for the element Ba in the halo star HD 140283. They found, specifically, 
that the fractional abundance of the odd isotopes

$f_{odd} = $[N($^{135}$Ba) + N($^{137}$Ba)]/N(Ba)

in this star was consistent with the solar system $r-$process isotopic ratio. 
Sm, like Ba, is an even-Z nucleus with one $p-$process isotope and six other 
stable ($s-$ and $r-$process admixed) isotopes. Thus (similarly to Ba), we can 
define for Sm 

$f_{odd} = $[N($^{147}$Sm) + N($^{149}$Sm)]/N(Sm)

For the pure $r$-process components of solar system isotopic abundances, we 
find that $f^{r}_{odd}$ = 0.36 for both the standard model (Burris et al. 
2000; Simmerer et al. 2004) and for the stellar model calculations given by 
Arlandini et al. (1999). For comparison, the Sm solar $s-$process values are 
$f^{s}_{odd}$ = 0.09 and $f^{s}_{odd}$ = 0.17 for the standard and stellar 
models, respectively. It might be possible to observe this $f^{r}_{odd 
}$isotopic mixture for Sm in a halo star, similarly to what Lambert {\&} 
Allende-Prieto did for Ba in HD 140283. Such an observation would provide a 
direct measure of the $r-$process isotopic contribution to the elemental Sm 
production in nucleosynthetic (e.g., supernovae) sites that were operating 
in the early Galaxy. Except for the one measurement of Ba and several Eu 
observations (Sneden et al. 2002, Aoki et al. 2003), there have been very 
few stellar isotopic abundance determinations. Sm (as noted above) has more 
of an $s-$process fraction than Eu, but less than Ba, in solar system material. 
Thus, any isotopic abundance observations of Sm would provide additional 
information about the early synthesis, and possible confirmation of the 
$r-$process only origin, for this element in halo stars. 

\begin{center}
8. SUMMARY 
\end{center}
We report here the largest-scale laboratory study to date of Sm II 
transition probablilities using modern methods. Specifically we have 
measured radiative lifetimes, accurate to $\pm $5{\%}, for 212 odd-parity 
levels of Sm \textsc{ii} using laser-induced fluorescence. The lifetimes are 
combined with branching fractions measured using Fourier-transform 
spectrometry to determine transition probabilities for 958 lines of Sm 
\textsc{ii}. This improved data set has been used to determine a new solar 
photospheric Sm abundance, log $\varepsilon $ = 1.00 $\pm $ 0.03, from 26 
lines. The spectra of three very metal-poor, neutron-capture-rich stars (CS 
22892-052, HD 115444 and BD+17$^{o}$3248) also have been reanalyzed, 
employing between 55 and 72 Sm II lines per star. We have compared the 
differences between the Sm abundances and the predicted solar system 
$r$-process only values in all three stars. We find with very small 
uncertainties that these ratios are the same. Utilizing additional recent 
experimental atomic data it was found that the abundance ratios of Sm 
relative to other rare earth elements in these stars are in agreeement, and 
are consistent with ratios expected from rapid neutron-capture 
nucleosynthesis (the $r-$process). The newly determined abundance values for Sm, 
based upon the much more precise atomic data, might possibly be employed to 
constrain predictions for the solar system abundances. Thus, the slight 
disagreement between the stellar Sm values and the predicted $r$-process only 
value might suggest a minor reassessment of the solar system $r$- (and $s$-) 
process contributions to Sm synthesis. We finally note the possibility of 
observing the isotopic mixture of Sm, similarly to what has been 
accomplished for Ba, in a metal-poor halo star. Such an observation would 
provide important additional information about the nature of nucleosynthesis 
early in the history of the Galaxy.

\begin{center}
ACKNOWLEDGMENTS
\end{center}

This work is supported by the National Science Foundation under grants 
AST-0205124 and 0506324 (JEL and EADH), AST{\-}0307495 (CS), and AST-0307279 
(JJC). J. E. Lawler is a guest observer at the National Solar Observatory 
and he is indebted to Mike Dulick and Detrick Branstron for help with the 1 
m Fourier transform spectrometer. Much of the solar/stellar abundance 
analysis was accomplished while CS was an Erskine Fellow at the University 
of Canterbury, Christchurch, New Zealand. Their Department of Physics and 
Astronomy is thanked for financial support and encouragement of this work. 
We are very grateful to Don Bord and J.-F. Wyart for sharing unpublished 
Cowan Code calculations with us.

\newpage 
\begin{center}
REFERENCES
\end{center}

Adams, D. L., {\&} Whaling, W. 1981, J. Opt. Soc. Am., 71, 1036

Anderson, H. M., Den Hartog, E. A., {\&} Lawler, J. E. 1996, J. Opt. Soc. 
Am. B{\bf , }13, 2382

Andersen, T., Poulsen, O., Ramanujam, P. S., {\&} Petkov, A. P. 1975, Solar 
Physics 44, 257

Aoki, W., Ryan, S. G., Iwamoto, N., Beers, T. C., Norris, J. E., Ando, H., 
Kajino, T., Mathews, G. J., Fujimoto, M. Y. 2003, ApJ, 582, L67

Arlandini, C., K\"{a}ppeler, F., Wisshak, K., Gallino, R., Lugaro, M., 
Busso, M., {\&} Straniero, O. 1999, ApJ, 525, 886

Bi\'{e}mont, E., Grevesse, N., Hannaford, P., {\&} Lowe, R. M. 1989, 
A{\&}Ap, 222, 307

Bi\'{e}mont, E., Baudoux, M., Kurucz, R. L., Ansbacher, W., {\&} Pinnington, 
E. H. 1991, A{\&}Ap, 249, 539

Bi\'{e}mont, E., Dutrieux, J.-F., Martin, I., Quinet, P. 1998, J. Phys. B: 
At. Molec. Opt. Phys., 31, 3321

Bi\'{e}mont, E., {\&} Quinet P. 2003, Physica Scripta, T105, 38

Blaise, J., Morillon, C., Schweighofer, M. G., {\&} Verges, J. 1969, 
Spectrochim. Acta B, 24, 405

Bord, D. J., Cowley, C. R., {\&} Mirijanian, D. 1998, Solar Physics, 178, 
221

Bord, D. J., {\&} Cowley, C. R. 2002, Solar Physics, 211, 3

Brault, J. W. 1976, J. Opt. Soc. Am., 66, 1081

Burris, D. L., Pilachowski, C. A., Armandroff, T. E., Sneden, C., Cowan, J. 
J., {\&} Roe, H. 2000, ApJ, 544, 302

Busso, M., Gallino, R., {\&} Wasserburg, G.J. 1999, ARA{\&}A, 37, 239

Cayrel, R., Depagne, E., Spite, M., Hill, V., Spite, F., Francois, P., Plez, 
B., Beers, T., Primas, F., Andersen, J., Barbuy, B., Bonifacio, P., Molaro, 
P., {\&} Nordstrom, B. 2004, A{\&}A, 416, 1117

Corliss, C. H., {\&} Bozman, W. R. 1962, {\it Experimental Transition Probabilities for Spectral Lines of Seventy Elements}, U. S. Natl. Bur. Standards 
Monograph 53, (Washington: U. S. Government Printing Office)

Cowan, J. J., Burris, D. L., Sneden, C., McWilliam, A., {\&} Preston, G. W. 
1995, ApJ, 439, L51

Cowan, J. J., Sneden, C., Burles, S., Ivans, I. I., Beers, T. C., Truran, J. 
W., Lawler, J. E., Primas, F., Fuller, G. M., Pfeiffer, B., {\&} Kratz, 
K.-L. 2002, ApJ,. 572, 861 

Cowan, J. J., {\&} Thielemann, F.-K. 2004, Phys. Today, 57, 47

Cowan, R. D. 1981, {\it The Theory of Atomic Structure and Spectra} (Berkeley, Univ. of California Press) 

Cowley, C. R. {\&} Corliss, C. H. 1983, MNRAS 203, 651

Cox, A. N. 2000, {\it Allen's Astrophysical Quantities} (New York: AIP Press) 

Curry, J. J., Den Hartog, E. A., {\&} Lawler, J. E. 1997, J. Opt. Soc. Am. 
B, 14, 2788

Danzmann, K., {\&} Kock, M. 1982, J. Opt. Soc. Am., 72, 1556

Den Hartog, E. A., Curry, J. J., Wickliffe, M. E., {\&} Lawler, J. E. 1998, 
Sol. Phys. 178, 239

Den~Hartog, E. A., Wiese, L. M. {\&} Lawler, J. E. 1999, J. Opt. Soc. Am. B, 
16, 2278

Den~Hartog, E. A., Fedchak, J. A., {\&} Lawler, J. E. 2001, J. Opt. Soc. Am. 
B 18, 861

Den~Hartog, E. A., Wickliffe, M. E., {\&} Lawler, J. E. 2002, ApJS, 141, 255

Den Hartog, E. A., Lawler, J. E., Sneden, C., {\&} Cowan, J. J. 2003, ApJS, 
148, 543

Den Hartog, E. A., Herd, M. T., Lawler, J. E., Sneden, C., Cowan, J. J., 
Beers, T. C. 2005, ApJ, 619, 639

Delbouille, L, Roland, G., {\&} Neven, L. 1973, {\it Photometric Atlas of the Solar Spectrum from lambda 3000 to lambda 10000}, (Li\`{e}ge, Inst. d'Ap., 
Univ. de Li\`{e}ge)

Derkatch, A., Ilyinsky, L., Mannervik, S., Norlin, L.-O., Rostohar, D., 
Royen, P., Schef, P., {\&} Bi\'{e}mont, E. 2002, Phys. Rev. A, 65, 062508

Edl\'{e}n, B. 1953, J. Opt. Soc. Am., 43, 339

Edl\'{e}n, B. 1966, Metrologia, 2, 71 

Fedchak, J. A., Den Hartog, E. A., Lawler, J. E., Palmeri, P., Quinet, P., 
{\&} Bi\'{e}mont, E. 2000, ApJ, 542, 1109

Gratton, R. G., {\&} Sneden, C. 1994, A{\&}Ap, 287, 927

Grevesse, N., {\&} Sauval, A. J. 1998, Space Sci. Rev., 85, 161

Grevesse, N., {\&} Sauval, A. J. 1999, A{\&}Ap, 347, 348

Grevesse, N., {\&} Sauval, A. J. 2002, Adv. Space. Res., 30, 3

Grigoriev, I. S., {\&} Melikhov, E. Z. 1997, {\it Handbook of Physical Quantities}, (Boca Raton, CRC Press) p. 
516

Guo, B., Ansbacher, W., Pinnington, E. H., Ji, Q., {\&} Berends, R. W. 1992, 
Phys. Rev. A, 46, 641

Hashiguchi, S., {\&} Hasikuni, M. 1985, J. Phys. Soc. Japan 54, 1290

Holweger, H., {\&} M\"{u}ller, E. A. 1974, Sol. Phys., 39, 19

Irwin, A. W. 1981, ApJS, 45, 621

Ivarsson, S., Litz\'{e}n, U., {\&} Wahlgren, G. M. 2001, Physica Scripta, 
64, 455

K\"{a}ppeler, F., Beer, H., {\&} Wisshak, K. 1989, Rep. Prog. Phys., 52, 945

Kono, A., {\&} Hattori, S. 1984, Phys. Rev. A, 29, 2981

Kurucz, R. L. 1998, in Fundamental Stellar Properties: {\it The Interaction between Observation and Theory}, IAU Symp. 189, ed 
T. R. Bedding, A. J. Booth and J. Davis (Dordrecht: Kluwer), p. 217

Lambert, D.L., {\&} Allende Prieto, C. 2002, MNRAS, 335, 325

Lawler, J. E., Bonvallet, G., {\&} Sneden, C. 2001a, ApJ, 556, 452

Lawler, J. E., Wickliffe, M. E., Cowley, C. R., {\&} Sneden, C. 2001b, ApJS, 
137, 341

Lawler, J. E., Wickliffe, M. E., Den Hartog, E. A., {\&} Sneden, C. 2001c, 
ApJ, 563, 1075

Lawler, J. E., Sneden, C., {\&} Cowan, J. J. 2004, ApJ, 604, 850

Li, Z. S., Svanberg, S., Quinet, P., Tordoir, X., {\&} Bi\'{e}mont, E. 1999, 
J. Phys. B: At. Molec. Opt. Phys., 32, 1731

Li, Z. S., {\&} Zhankui J. 1999, Physica Scripta, 60, 414

Lodders, K. 2003, ApJ, 591, 1220

Martin, W.C., Zalubas, R., {\&} Hagan, L. 1978,{\it  Atomic Energy Levels {\-} The Rare Earth Elements}, NSRDS{\-}NBS 60 
(Washington: U. S. G. P. O.) p. 174

Martin, W. C., Sugar, J., {\&} Musgrove, A. 2000, NIST Atomic Spectra 
Database, (http://physics.nist.gov/cgi{\-}bin/AtData/main{\_}asd)

Masterman D., Rosner, S. D., Scholl, T. J., Sharikova, A., {\&} Holt, R. A. 
2003, Can J. Phys., 81, 1389

McWilliam, A., Preston, G. W., Sneden, C., {\&} Searle, L. 1995, ApJ, 109, 
2757

Meggers, W. F., Corliss, C. H., and Scribner, B. F. 1961, {\it Tables of Spectral Line Intensities}, U. S. Natl. Bur. 
Standards Monograph 32, (Washington: U.S. G.P.O.)

Meggers, W. F., Corliss, C. H., and Scribner, B. F. 1975, {\it Tables of Spectral Line Intensities}, U. S. Natl. Bur. 
Standards Monograph 145, (Washington: U.S. G. P. O.)

Moore, C. E., Minnaert, M. G. J., {\&} Houtgast, J. 1966, {\it The Solar Spectrum 2934 {\AA} to 8770 {\AA}}, NBS Monograph 61 
(Washington: U.S. G. P. O.)

Obbarius H. U., {\&} Kock, M. 1982, J. Phys. B: At. Mol. Opt. Phys., 15, 527

Palmeri, P., Quinet, P., Wyart, J.-F., {\&} Bi\'{e}mont, E. 2000, Physica 
Scripta, 61, 323

Pinnington, E. H., Rieger, G., {\&} Kernahan, J. A. 1997, Phys. Rev. A, 56, 
2421

Quinet, P., Palmeri, P., {\&} Bi\'{e}mont, E. 1999a, J. Quant. Spectrosc. 
Radiat. Transfer, 62 625

Quinet, P., Palmeri, P., Bi\'{e}mont, E., McCurdy, M. M., Rieger, G., 
Pinnington, E. H., Wickliffe, M. E., {\&} Lawler, J. E. 1999b, MNRAS, 307, 
934

Rieger, G., McCurdy, M. M., {\&} Pinnington E. H. 1999, Phys. Rev. A, 60, 
4150

Rostohar, D., Andersson, K., Derkatch, A., Hartman, H., Mannervik, S., 
Norlin, L.-O., Royen, P., Schmidtt, A., Tordoir, X. 2001, Physica Scripta, 
64, 237 

Ryan, S. G., Norris, J. E., and Beers, T. C. 1996, ApJ, 471, 254

Saffman L., {\&} Whaling W. 1979, J. Quant. Spectrosc. Radiat. Transfer, 21, 
93

Scholl, T. J., Holt, R. A., Masterman, D., Rivest, R. C., Rosner, S. D., 
{\&} Sharikova, A. 2002, Can. J. Phys., 80, 1621

Simmerer, J., Sneden, C., Cowan, J. J., Collier, J., Woolf, V. M., {\&} 
Lawler, J. E. 2004, ApJ, 617, 1091 

Sneden, C. 1973, ApJ, 184, 839

Sneden, C., McWilliam, A., Preston, G. W., Cowan, J. J., Burris, D. L., and 
Armosky, B. J. 1996, ApJ,  467, 819

Sneden, C., Cowan, J. J., Lawler, J. E., Burles, S., Beers, T. C., Fuller, 
G. M. 2002, 566, L25

Sneden, C., Cowan, J. J., Lawler, J. E., Ivans, I. I., Burles, S., Beers, T. 
C., Primas, F., Hill, V., Truran, J. W., Fuller, G. M., Pfeiffer, B., {\&} 
Kratz, K.-L. 2003, ApJ, 591, 936

Sneden, C., {\&} Cowan, J. J. 2003, Science, 299, 70

Taylor, P., Roberts, M., Gateva-Kostova, S. V., Clarke, R. B. M., Barwood, 
G. P., Rowley, W. R. C., {\&} Gill, P. 1997, Phys. Rev. A, 56, 2699

Truran, J. W., Cowan, J. J., Pilachowski, C. A., {\&} Sneden, C. 2002, PASP, 
114, 1293

Vogel, O., Edvardsson, B., W\"{a}nnstr\"{o}m, A., Arneson, A., {\&} Hallin, 
R. 1988, Physica Scripta, 38, 567

Volz, U., {\&} Schmoranzer, H. 1998, in AIP Conf. Proc. 434, {\it Atomic and Molecular Data and Their Applications}, ed. P. J. 
Mohr and W. L. Wiese (Woodbury, NY:AIP), p. 67

Ward, L. 1985, Mon. Not. R. Astr. Soc. 213, 17

Weiss, A. W. 1995, Phys. Rev. A, 51, 1067

Westin, J., Sneden, C., Gustafsson, B., {\&} Cowan, J.J. 2000, ApJ, 530, 783

Whaling, W., Carle, M. T., {\&} Pitt, M. L. 1993, J. Quant. Spectrosc. 
Radiat. Transfer 50, 7

Wickliffe, M. E., {\&} Lawler, J. E. 1997, J. Opt. Soc. Am. B, 14, 737

Wickliffe, M. E., Lawler, J. E., {\&} Nave, G. 2000, J. Quant. Spectrosc. 
Radiat. Transfer, 66, 363

Wyart, J.-F., {\&} and Bauche-Arnoult, Cl. 1981, Physica Scripta, 22, 583

Xu, H. L., Jiang, Z. K., {\&} Svanberg, S. 2003a, J. Phys. B: At. Molec. 
Opt. Phys., 36, 411

Xu, H. L., Svanberg, S., Quinet, P., Garnir, H. P., {\&} Bi\'{e}mont, E. 
2003b, J. Phys. B: At. Molec. Opt. Phys., 36, 4773

Yan, Z-C, Tambasco, M., {\&} Drake, G. W. F. 1998, Phys. Rev. A, 57, 1652

Youssef, N. H., {\&} Khalil, N. M. 1989, A{\&}Ap, 215, 165

Yu, N., {\&} Maleki L. 2000, Phys. Rev. A, 61, 022507

Zhang, Z. G., Li Z. S., Lundberg, H., Zhang, K. Y., Dai, Z. W., Jiang, Z. 
K., Svanberg, S. 2000, J. Phys. B: At. Molec. Opt. Phys., 33, 521

Zhang, Z. G., Persson A., Li, Z. S., Svanberg, S., {\&} Zhankui, J. 2001a, 
Eur. J. Phys. D, 13, 301

Zhang, Z. G., Svanberg, S., Jiang, Z. K., Palmeri, P., Quinet, P., {\&} 
Bi\'{e}mont, E. 2001b, Physica Scripta, 63, 122

Zhao, R. C., Huang, W., Xu, X. Y., Tong, X. M., Qu, Y. Z., Xu, C. B., {\&} 
Xue, P. 1996, Phys. Rev. A, 53, 3994

Zhiguo, Z., Zhongshan, L., {\&} Zhankui J. 1999, Eur. Phys. J. D, 7, 499

\newpage 
\begin{center}
FIGURE CAPTIONS
\end{center}

Figure 1. Partial Grotrian diagram for singly ionized samarium. 

Figure 2. Comparison of the log({\it gf}) values from Xu et al. (2003b) to log({\it gf}) 
values from this paper as a function of wavelength. Solid round symbols are 
for lines which Xu et al. gave a log({\it gf}) uncertainty $<$ 0.04 dex, open circle 
symbols are for lines which Xu et al. gave a log({\it gf}) uncertainty of $<$ 0.11 
dex, and the ``x'' symbols are used for lines which Xu et al. gave a 
log({\it gf}) uncertainty $>$ 0.11 dex.

Figure 3. Comparison of the log({\it gf}) values from Xu et al. (2003b) to log({\it gf}) 
values from this paper as a function of our log({\it gf}) value. The symbols have 
the same meaning as in Figure 2.

Figure 4. Comparison of the log({\it gf}) values from Xu et al. (2003b) to log({\it gf}) 
values from this paper as a function of upper level energy E$_{upper}$. The 
symbols have the same meaning as in Figure 2.

Figure 5. Comparison of the log({\it gf}) values from Saffman {\&} Whaling (1979), 
normalized to our experimental radiative lifetimes, to log({\it gf}) values from 
this paper as a function of wavelength. 

Figure 6. Comparison of the log({\it gf}) values from Saffman {\&} Whaling (1979), 
normalized to our experimental radiative lifetimes, to log({\it gf}) values from 
this paper as a function of our log({\it gf}) value. 

Figure 7. Comparison of the log({\it gf}) values from Saffman {\&} Whaling (1979), 
normalized to our experimental radiative lifetimes, to log({\it gf}) values from 
this paper as a function of upper level energy.

Figure 8. Comparison of radiative lifetimes $\tau _{other}$ measured using 
LIF by: Bi\'{e}mont et al. (1989) solid round symbol, Scholl et al. (2002) 
open circle symbol, Vogel et al. (1988) ``x'' symbol, and Xu et al. (2003b) 
``+'' symbol to radiative lifetimes $\tau _{this\_expt}$ measured in this 
experiment. The ordinate scale is designed to match the ordinate scale of 
Figures 2-7.

Figure 9. Same as Figure 8 except with an expanded ordinate.

Figure 10. Relative strength factors, defined as log({\it gf}) - {\it $\theta  
\chi $}, for Sm\textsc{ 
ii} transitions. Reduced widths of weak lines should be proportional to 
these factors. For these computations, {\it $\theta $ } = 1.0, the mean of inverse effective 
temperature of the four program stars. The 4424.35 {\AA} is specially noted 
in the plot, as it is the strongest yet reasonably unblended Sm\textsc{ ii} 
feature in our solar and stellar spectra. The text discusses this line and 
the four "strongest lines" that are noted in this figure with a large 
enclosing circle.

Figure 11. Individual Sm\textsc{ ii} line abundances in the solar 
photosphere derived in this work (upper panel) and in four previous studies 
(lower panel). In the upper panel, a dotted horizontal line denotes the mean 
abundance. In the lower panel, symbols as indicated in the plot denote the 
results of Andersen et al. (1975: And75 in the plot legend), Saffman {\&} 
Whaling (1979: Saf79), Vogel et al. (1988: Vog88), Youssef {\&} Khalil 
(1989: You89), and Bi\'{e}mont et al. (1989: Bie89).

Figure 12. Abundances of Sm in the very metal-poor, $n-$capture-rich giant star 
BD+17$^{o}$3248. The upper panel contains our published results (Cowan et 
al. 2002), and the lower panel shows the data of this paper.

Figure 13. The neutron-capture elemental abundance pattern in the Galactic 
halo stars CS 22892-052, HD 115444, and BD+17$^{o}$3248 compared with the 
(scaled) Solar System $r-$process abundances (solid line) and the total Solar 
System meteoritic abundances recommended by Lodders (2003; dashed line). The 
abundances of all of the stars have been compared by normalizing the 
abundance distributions at the $r-$process element Eu. In the top panel, stellar 
abundances are those reported in the original papers on these stars (HD 
115444, Westin et al. 2000; BD+17$^{o}$3248, Cowan et al. 2002; CS 
22892-052, Sneden et al. 2003). In the bottom panel the comparisions are 
repeated for these stars, but substituting in the abundances of Nd, Ho, and 
Sm derived in the most recent papers of this series (see text). Also shown 
are solar photospheric abundances, with values of La, Nd, Sm, Eu, Tb, and Ho 
taken from this series, otherwise from Lodders (2003).

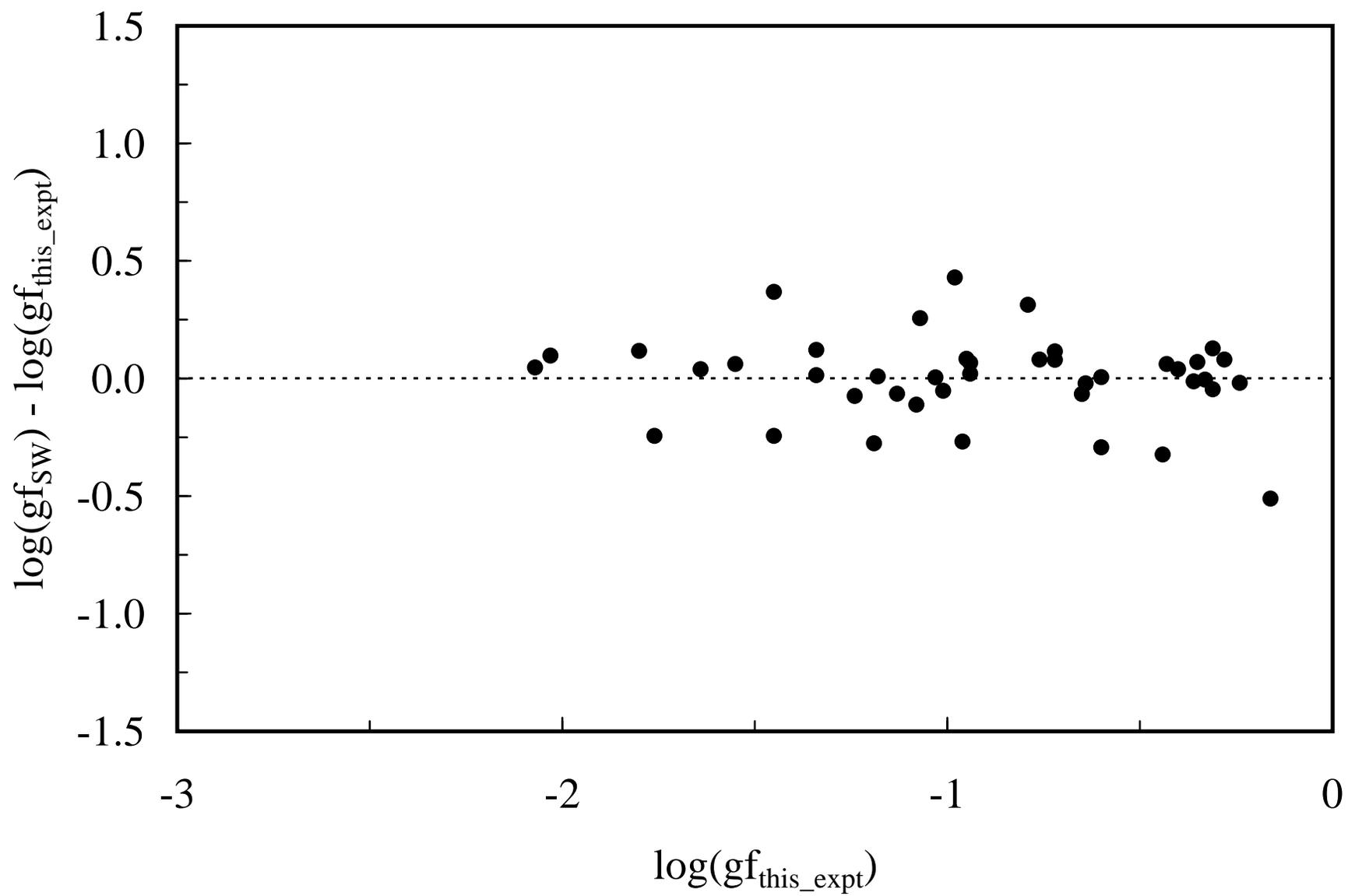
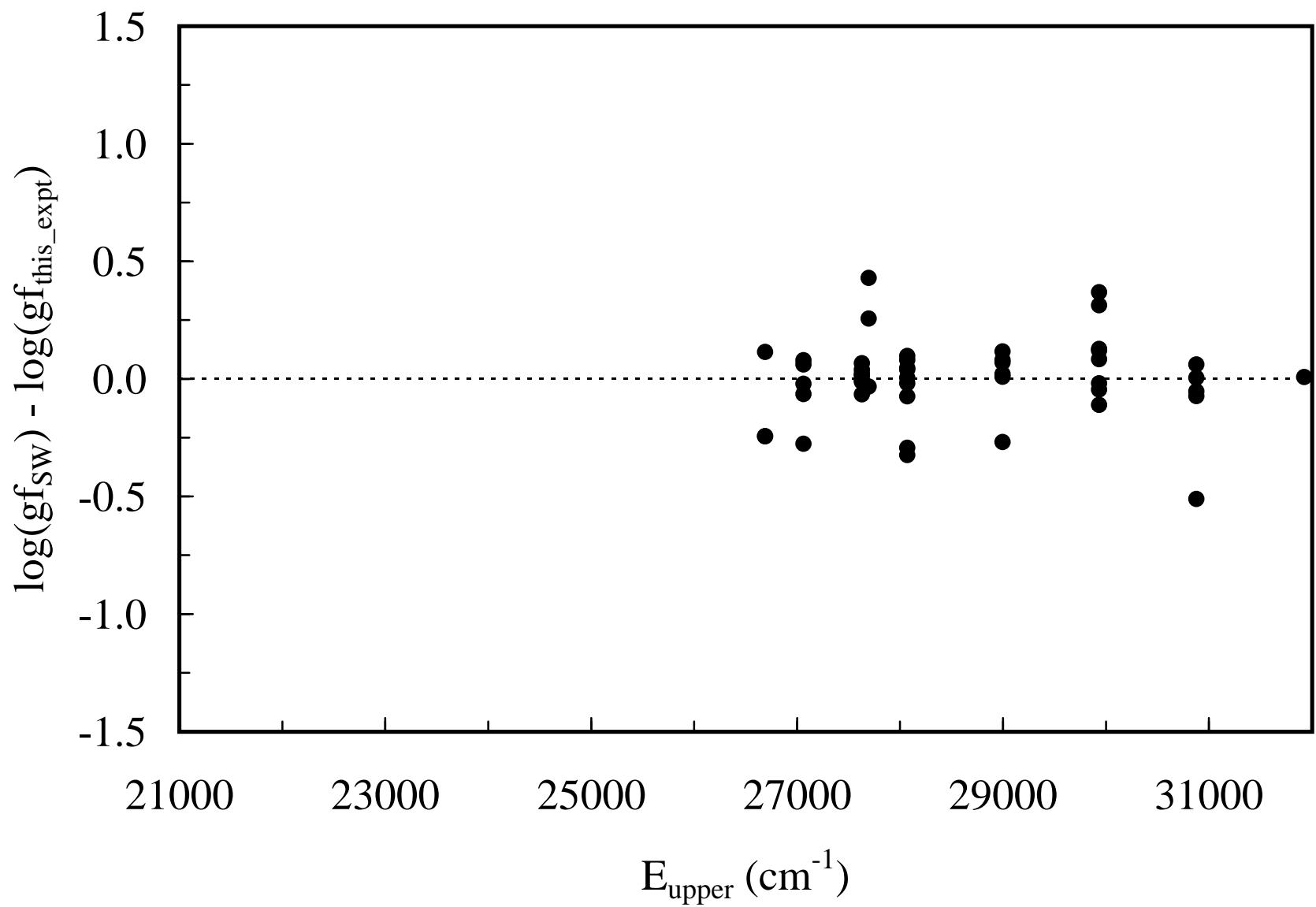
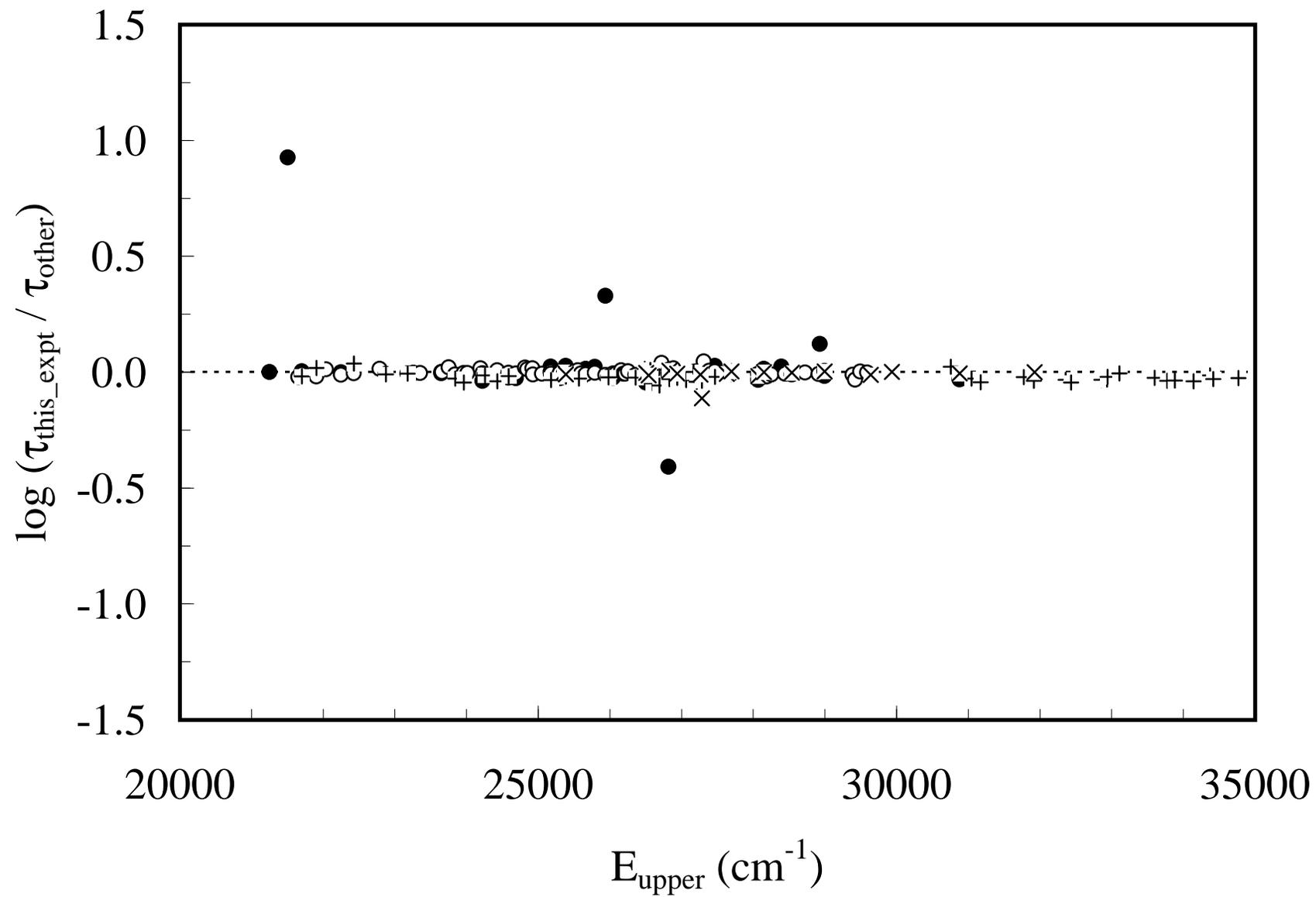
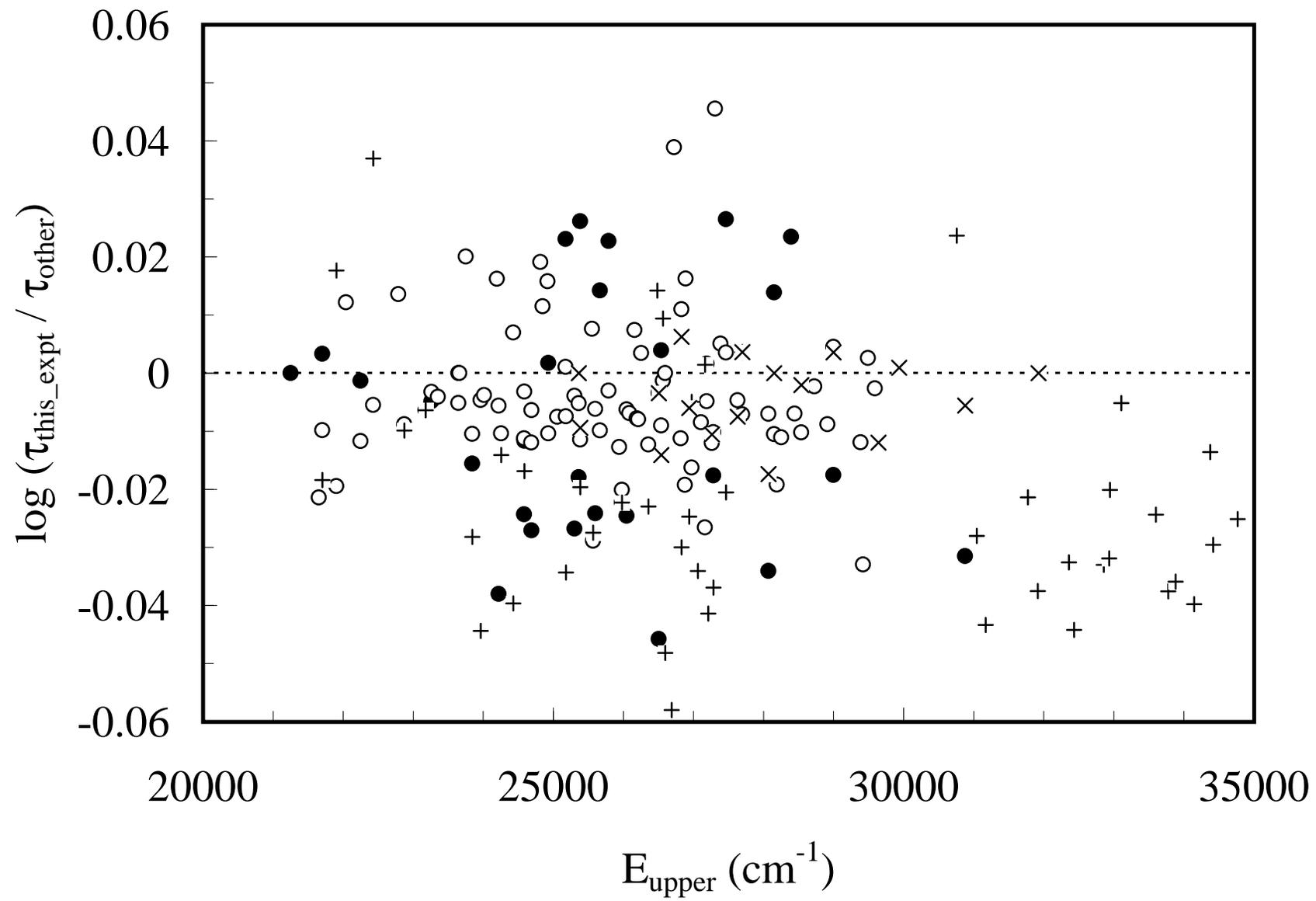
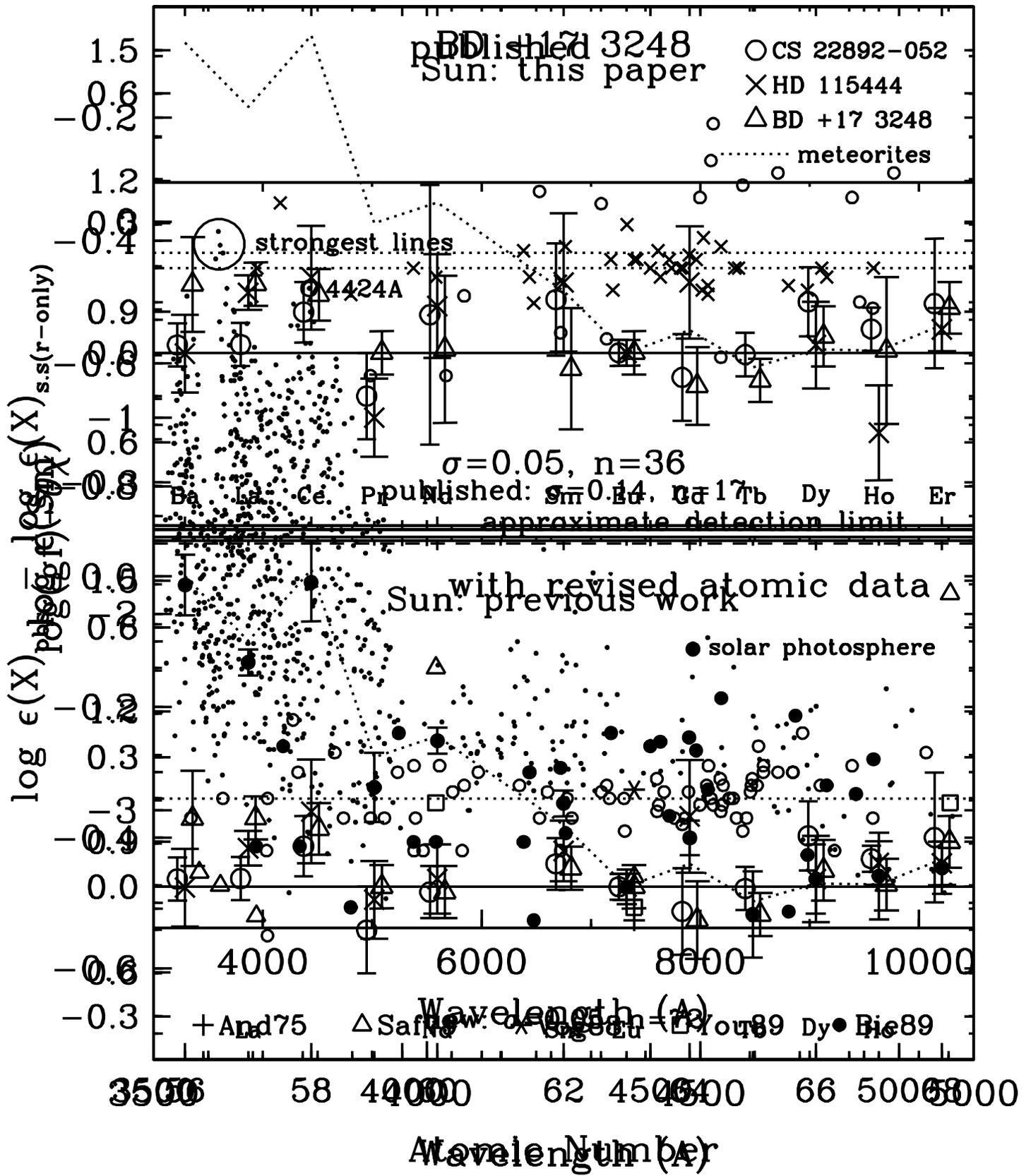
\end{document}